\documentclass[a4paper, 12pt]{article}
\usepackage[a4paper,margin=25mm]{geometry}
\usepackage[pdftex]{graphicx}
\usepackage{authblk}
\usepackage{amsmath}
\usepackage{amssymb}
\usepackage{amsfonts}
\usepackage{mathrsfs}
\usepackage{bbm}
\usepackage{listings}
\usepackage[pdftex]{color}
\usepackage{here}
\usepackage{algorithm, algorithmic}

\definecolor{codegreen}{rgb}{0,0.6,0}
\definecolor{codegray}{rgb}{0.5,0.5,0.5}
\definecolor{codepurple}{rgb}{0.58,0,0.82}
\definecolor{codefunc}{rgb}{0.85, 0.18, 0.50}

\lstdefinelanguage{PythonFuncColor}{
  language=Python,
  morekeywords=[2]{where,argsort,array,concatenate,full\_like,at,maximum,minimum,stop\_gradient,matmul,softmax,gumbel\_softmax\_sample,asarray},
}

\lstdefinestyle{mystyle}{
    language=PythonFuncColor,
    backgroundcolor=\color{white},   
    commentstyle=\color{codegreen},
    keywordstyle=\color{blue},
    keywordstyle=[2]\color{codefunc},
    breakatwhitespace=false,        
    breaklines=true,                 
    captionpos=b,                    
    keepspaces=true,                 
    showspaces=false,                
    showstringspaces=false,
    showtabs=false,                  
    tabsize=1,
    basicstyle=\footnotesize\ttfamily,
}
\makeatother

\title{Ultra-fast Traffic Nowcasting and Control via Differentiable Agent-based Simulation}

\author[1]{Fumiyasu Makinoshima\thanks{Corresponding author: f.makinoshima@fujitsu.com}}
\author[1]{Yuya Yamaguchi} 
\author[1]{Eigo Segawa}
\author[2]{Koichiro Niinuma}
\author[3,4]{Sean Qian}

\affil[1]{Fujitsu Limited, Kawasaki, Japan}
\affil[2]{Fujitsu Research of America, Pittsburgh, PA, USA}
\affil[3]{Department of Civil and Environmental Engineering, Carnegie Mellon University, Pittsburgh, PA, USA}
\affil[4]{Heinz College of Information Systems and Public Policy, Carnegie Mellon University, Pittsburgh, PA, USA}

\date{}

\begin{document}
\maketitle

\begin{abstract}
Traffic digital twins, which inform policymakers of effective interventions based on large-scale, high-fidelity computational models calibrated to real-world traffic, hold promise for addressing societal challenges in our rapidly urbanizing world.
However, conventional fine-grained traffic simulations are non-differentiable and typically rely on inefficient gradient-free optimization, making calibration for real-world applications computationally infeasible.
Here we present a differentiable agent-based traffic simulator that enables ultra-fast model calibration, traffic nowcasting, and control on large-scale networks.
We develop several differentiable computing techniques for simulating individual vehicle movements, including stochastic decision-making and inter-agent interactions, while ensuring that entire simulation trajectories remain end-to-end differentiable for efficient gradient-based optimization.
On the large-scale Chicago road network, with over 10{,}000 calibration parameters, our model simulates more than one million vehicles at 173 times real-time speed.
This ultra-fast simulation, together with efficient gradient-based optimization, enables us to complete model calibration using the previous 30 minutes of traffic data in 455~s, provide a one-hour-ahead traffic nowcast in 21~s, and solve the resulting traffic control problem in 728~s.
This yields a full calibration--nowcast--control loop in under 20 minutes, leaving about 40 minutes of lead time for implementing interventions.
Our work thus provides a practical computational basis for realizing traffic digital twins. 
\end{abstract}

\section*{Introduction}
Rapid global urbanization represents a pivotal moment in human history, with urban dwellers now constituting the nearly half of the world's 8.2 billion population~\cite{un_2025}.
The rapid urbanization has elevated the traffic congestion management to a critical global challenge, as it has significant effects on economic productivity~\cite{sweet_2014,fattah_2022} and environmental and public health~\cite{lelieveld_2015,knittel_2016}, beyond mere inconvenience.
Given the difficulty and potentially irreversible impacts of implementing traffic policies in real-world, traffic simulation has become an indispensable tool for identifying effective strategies within a virtual space in computers.
This has led to the development of various simulation models~\cite{fellendorf_2010,horni_2016,lopez_2018} simulating vehicle movements at a fine scale with micro or mesoscopic models, which are now widely applied to inform practical policy-making.  

While simulation is a powerful tool for policy-making, its real-world effectiveness hinges on calibration, the process of aligning the model with empirical data.
A well-calibrated simulator enables the exploration of reliable and effective strategies~\cite{barcelo_2010}.
This calibration process, however, becomes a critical bottleneck for fine-grained traffic simulation models.
Because of their inherent stochasticity and complex vehicle interactions, such models are non-differentiable and must typically be calibrated using gradient-free optimization algorithms~\cite{ma_2002,shultz_2004,park_2005,ma_2007,lee_2009}, which are notoriously inefficient and require vast numbers of simulation runs.
As the spatial scale of the simulation expands, the number of calibration parameters grows rapidly, further compounding the problem and making the optimization computationally intractable.
As a result, the practical application of calibrated microscopic models remains confined to small-scale scenarios with only a few dozen parameters~\cite{hollander_2008}, falling short of the needs for addressing complex, large-scale urban systems.
On the other hand, real-time traffic observation data for calibration is now becoming readily available from various observation channels~\cite{zhu_2019}.
This situation underscores the urgent need for advanced simulation technologies capable of translating this wealth of data into actionable insights for timely transportation policy making.

The emerging paradigm of differentiable simulation presents a promising pathway to overcome this challenge.
Effectively using automatic differentiation developed mainly for machine learning~\cite{baydin_2017}, the differentiable simulations calculate the dynamics of the equation-based white-box models while enabling the access to the gradient information of the simulated process and efficient gradient-based optimization~\cite{newbury_2024}.
Differentiable simulation techniques first demonstrated their effectiveness for rapid parameter calibration in the field of physical simulation~\cite{degrave_2019,hu_2020,bezgin_2023}.
More recently, their application has been extended to social simulations~\cite{andelfinger_2021,andelfinger_2023,chopra_2023,dyer_2023,querabofarull_2025}, where comparable benefits have been reported.
In traffic engineering, while efforts to apply this paradigm to parameter calibration are emerging, they have been largely confined to macroscopic, continuous models~\cite{wu_2018,ma_2020,patwary_2023,guarda_2024,kim_2024,zhou_2025}.
The few pioneering works on microscopic models~\cite{andelfinger_2021,andelfinger_2023,son_2022,son_2025} have come at the cost of simplification, often omitting complex behaviors such as agents' stochastic decision-making and their interactions with other agents.
The core challenge lies in consistently differentiating stochastic choice of vehicles, vehicle transfers between road links, and vehicle-to-vehicle interactions.
Consequently, developing a differentiable fine-grained traffic simulator with the fidelity to capture traffic dynamics in complex, real-world networks has yet to be achieved.

Here we introduce an end-to-end differentiable, agent-based model that simulates the dynamics of over one million vehicles in real-time.
Its design is grounded in traffic flow theory and employs novel computational techniques to maintain full differentiability throughout the simulation, even with complex stochastic behaviors of vehicles.
The model is fully tensorized to leverage massive parallelization on hardware accelerators, achieving simulation speed significantly exceeding real-time.
We demonstrate its capability on a large-scale, real-world road network with over 10{,}000 calibration parameters.
This exceptional speed, combined with efficient gradient-based optimization, enables model calibration, nowcasting, and intervention planning to be completed several times faster than real time, so that the full pipeline finishes with ample lead time for implementing interventions, thereby realizing a practical, operational traffic digital twin for a city.
By bridging the gap between traditional simulators and the vast amount of available traffic data, our work paves the way for next-generation intelligent transportation systems to address pressing societal challenges.

\section*{Results}
\subsection*{Differentiable agent-based traffic simulation}
\begin{figure}[H]
  \centering
  \includegraphics[width=\textwidth]{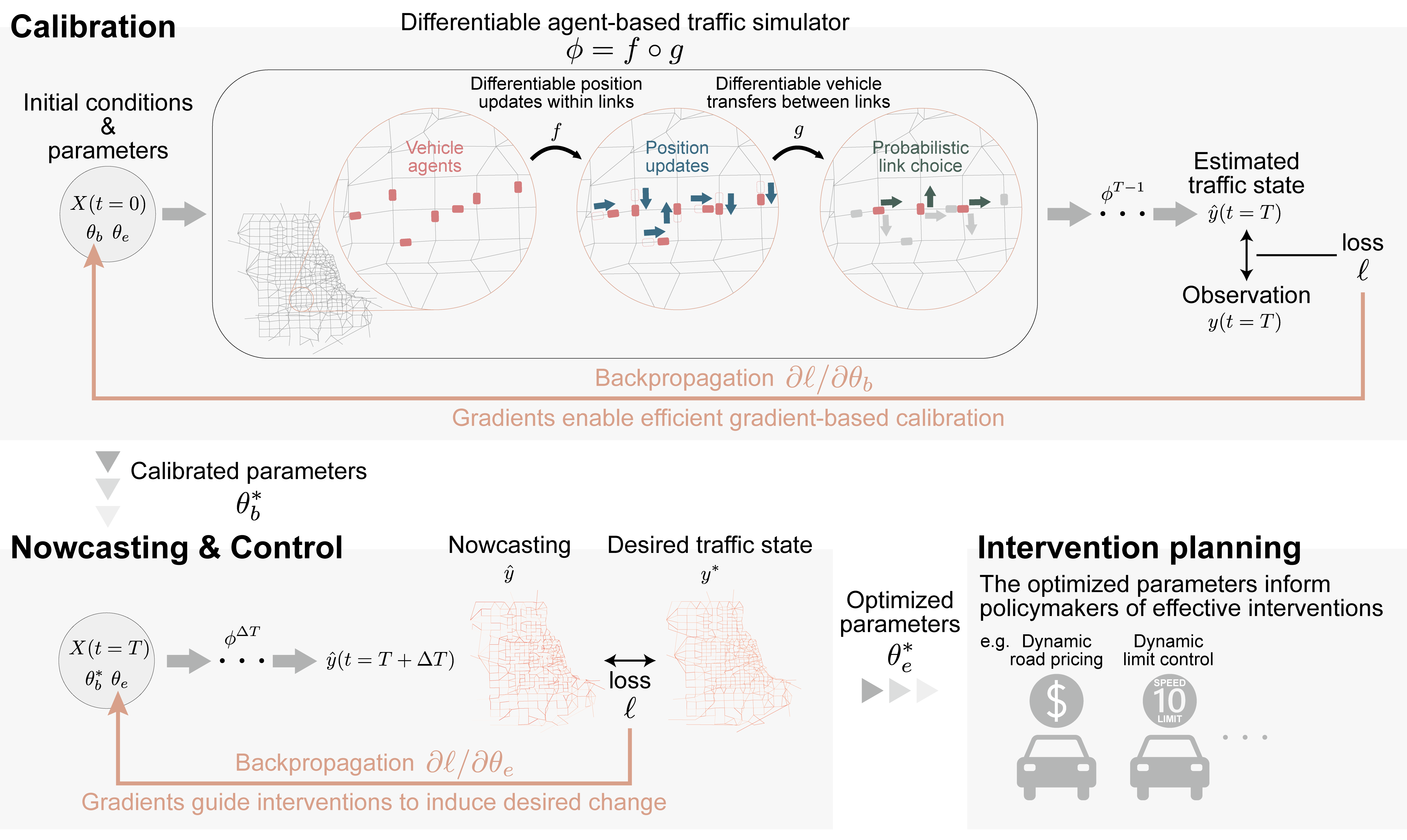}
  \caption{Framework of the differentiable agent-based traffic simulation for traffic nowcasting and control. The framework consists of two main stages: calibration and nowcasting/control.
           (1) Calibration: The simulation runs with initial conditions $X(t=0)$, optimizable behavioral parameters ($\theta_b$), and fixed environmental parameters ($\theta_e$).
           By minimizing the loss $\ell$ between the simulated traffic state $\hat{y}(t=T)$ and ground-truth observations $y(t=T)$ via gradient-based optimization using the gradient ${\partial}{\ell}/{\partial}{\theta_b}$, the behavioral parameter is calibrated to $\theta^{*}_b$, accurately replicating observed traffic dynamics.
           (2) Nowcasting and Control: Using the calibrated parameter $\theta^{*}_b$, the current traffic state $X(t=T)$, and the current environmental parameter $\theta_e$, the model performs nowcasting to predict the near-future state $\hat{y}(t=T+{\Delta}T)$.
           To control traffic, a desired state $y^{*}$ is set.
           The gradient with respect to the environmental parameter, ${\partial}{\ell}/{\partial}{\theta_e}$, is then calculated to determine the optimal interventions $\theta^{*}_e$ (e.g., dynamic pricing, dynamic speed limit control) that guide the system towards this desired state.
           }
  \label{fig:schematic}
\end{figure}

We developed an end-to-end differentiable agent-based traffic simulator for real-time traffic nowcasting and control (see Methods and Supplementary Information A--D for details).
The overall framework is illustrated in Fig.~\ref{fig:schematic}.
Given the initial traffic state and two types of parameters, namely behavioral model parameters $\theta_b$ and environmental parameters $\theta_e$, the simulator computes the traffic state over a time horizon of length $T$.
While the environmental parameters, such as road pricing, can be specified objectively and, in some cases, decided by policy makers, the behavioral parameters governing how travelers perceive the environment and make decisions are not directly observable and must be inferred from data.

A single forward step of our simulator, denoted by $\phi$, can be viewed as the composition of a position update function $f$ and a vehicle transfer function $g$.
The position update function $f$ advances vehicle positions over one time step based on traffic flow theory, whereas the vehicle transfer function $g$ governs probabilistic next-link choices and stochastic merging behavior.
In this study, we construct differentiable versions of these processes while still simulating vehicle movements in a manner that is faithful to traffic flow theory, without ad-hoc simplifications of inter-agent interactions or stochasticity (see Methods for details).
Because these processes are differentiable, the composite function $\phi$ and the simulated trajectories $\phi^{T}$ over $T$ time steps are end-to-end differentiable.
Consequently, the gradient of a loss function, which quantifies the discrepancy between the simulated traffic state $\hat{y}$ and the observed state $y$, with respect to the behavioral parameters $\theta_b$, can be computed efficiently via automatic differentiation, enabling scalable gradient-based calibration of the simulation model.

After calibration, we nowcast the traffic state at time $T + {\Delta}T$ by simulating the system with the calibrated behavioral parameters $\theta^{*}_b$ and the environmental parameters $\theta_e$.
If the forecasted traffic state exhibits undesirable features, such as severe congestion on particular roads, we define the deviation between the forecasted and a desired traffic state (e.g., a less congested state) as a loss function.
The gradient of this loss with respect to the environmental parameters $\theta_e$ can then be obtained via automatic differentiation, which effectively guides the system towards the desired state.
The resulting optimal parameters $\theta^{*}_e$ inform policy makers about effective interventions, such as road pricing strategies.
If these steps, that is, calibration, nowcasting, and solution of the associated control problems, can be performed within a very short time period, they pave the way towards a digital twin of urban traffic that continuously updates the computational model using real-world traffic observations and guides interventions to improve real-world traffic conditions.
In the subsequent sections, we demonstrate that, with the efficient gradient-based optimization enabled by automatic differentiation, these complex tasks, namely calibration, traffic nowcasting, and control, can be executed faster than real time, even for city-scale networks.

\subsection*{Gradient-based model calibration}
\begin{figure}[H]
  \centering
  \includegraphics[width=\textwidth]{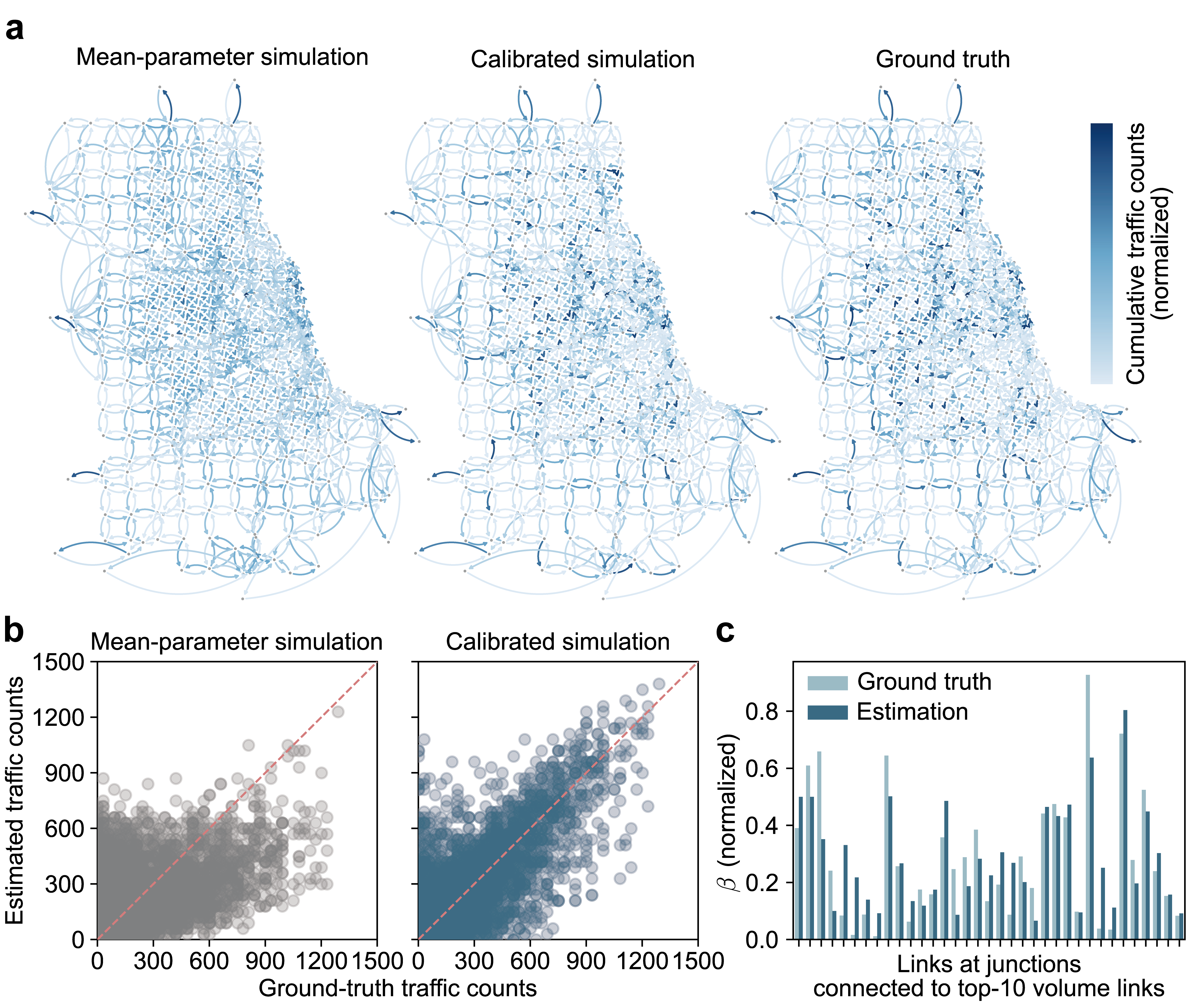}
  \caption{Calibration results on the Chicago Sketch network.
          (a) Cumulative traffic counts from the mean-parameter simulation and the calibrated simulation, compared against ground-truth counts.
          (b) Five-minute traffic counts from the mean-parameter simulation and the calibrated simulation, compared against ground-truth counts.
          (c) Estimated $\beta$ values compared against ground-truth $\beta$ values. Because the choice probability of the next links is determined only by differences in utility, $\beta$ is normalized for comparison. For clarity, we display only the $\beta$ values of links at junctions connected to the top-10 links by traffic volume.
          }
  \label{fig:chicago_calibration}
\end{figure}

As we verified that the calibration can be substantially accelerated with a larger platoon size $\Delta n$ without adverse effects on the loss (see Supplementary Information~E), we applied the developed model to the large-scale Chicago network, involving 1{,}000{,}020 vehicles and 10{,}284 calibration parameters, with $\Delta n = 30$ (see Methods).
Whereas the simulation with mean parameters failed to reproduce the spatially heterogeneous traffic patterns of the ground truth, the simulation with the gradient-based calibration successfully reproduced the complex traffic patterns (Fig.~\ref{fig:chicago_calibration}a).
Quantitative comparisons against ground-truth traffic counts (every 5 min) showed that the calibrated simulation provides a good approximation of the true traffic state (mean absolute error (MAE) of 52.6; Pearson's $r = 0.83$), whereas the mean-parameter simulation failed to capture the actual traffic trend (MAE of 87.4; Pearson's $r = 0.60$), yielding a 40\% improvement in accuracy (Fig.~\ref{fig:chicago_calibration}b).
The estimated behavioral model parameter $\beta$ reproduced the ground-truth parameter distribution (Fig.~\ref{fig:chicago_calibration}c), indicating that the model successfully estimated the behavioral model from noisy, partial, and aggregate traffic counts and provided a good approximation of the actual traffic states.
The forward simulation and its automatic differentiation are quite fast, and this large-scale gradient-based calibration required only 455.3~s, which corresponds to 4.0$\times$ real-time.

\subsection*{Real-time traffic nowcasting}
\begin{figure}[H]
  \centering
  \includegraphics[width=\textwidth]{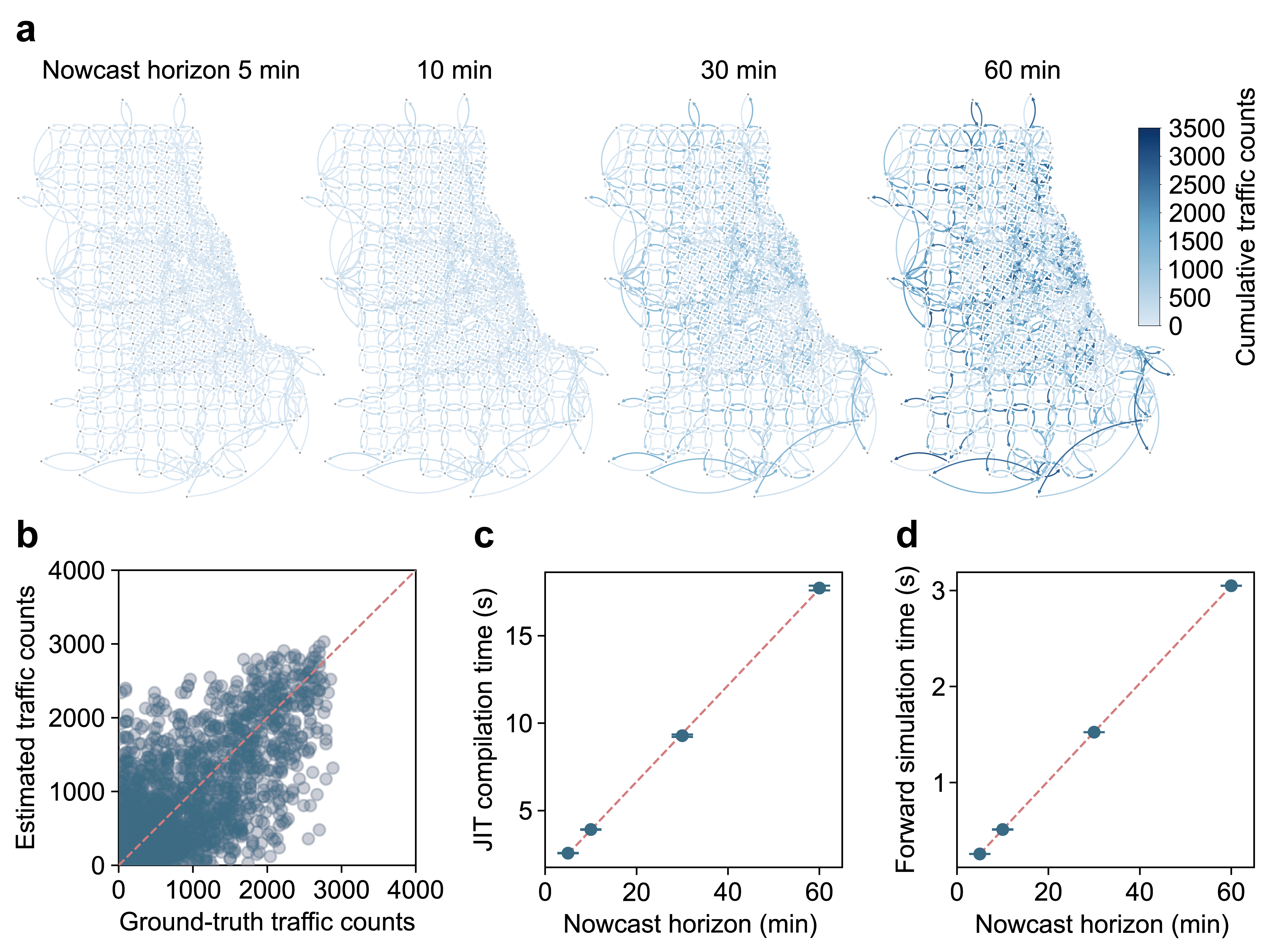}
  \caption{Traffic nowcasting results on the Chicago Sketch network.
          (a) Estimated cumulative traffic counts at nowcast horizons of 5, 10, 30, and 60 minutes from the current state.
          (b) Estimated cumulative traffic counts on links at the 60-minute nowcast horizon compared against ground-truth counts.
          (c) JIT compilation time for different nowcast horizons.
          (d) Forward simulation time for different nowcast horizons.
          In panels c and d, whiskers indicate the standard deviation across five trials.
          The dashed lines represent the regression results, indicating the linear scalability of the nowcasting computation.
    }
  \label{fig:chicago_nowcasting}
\end{figure}

After the calibration process, we ran forward simulations to nowcast future traffic states using the calibrated parameters (Fig.~\ref{fig:chicago_nowcasting}).
By evolving the system from the current state with the calibrated parameters, we obtained future traffic states conditioned on the observed traffic flows during the previous 30 minutes (Fig.~\ref{fig:chicago_nowcasting}a).
Because only partial and noisy traffic counts are observed and agents' route choices are stochastic, the calibrated model exhibits estimation errors that grow with the nowcast horizon.
In this nowcasting test, we observed a systematic bias of approximately 6.9 veh/min per link on average, which led to a discrepancy between the estimated and ground-truth cumulative traffic counts after 60 minutes (Fig.~\ref{fig:chicago_nowcasting}b).
Nonetheless, the nowcast captures the overall pattern of actual traffic flows (Pearson's $r = 0.69$), which can still be informative for practical intervention planning.
The nowcasts can be computed much faster than real time, and the computational cost increases approximately linearly with the nowcast horizon (Fig.~\ref{fig:chicago_nowcasting}c, d), because the length of the time-integration loop to be compiled and simulated increases in proportion to the nowcast horizon.
For a one-hour-ahead nowcast, the just-in-time (JIT) compilation and forward simulation required 17.7~s and 3.0~s, respectively, corresponding to an overall speed of 173$\times$ real time for the nowcasting process.

\subsection*{Gradient-guided traffic control}
\begin{figure}[H]
  \centering
  \includegraphics[width=\textwidth]{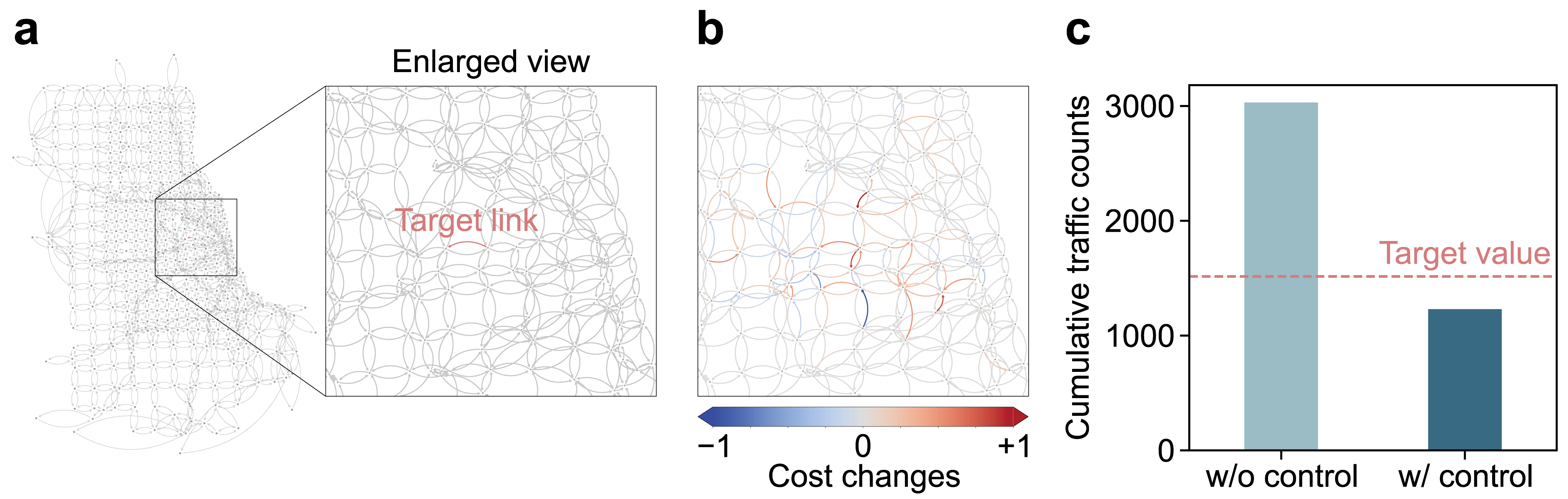}
  \caption{Traffic control results on the Chicago Sketch network.
          (a) Network view highlighting the target link to be controlled via cost changes.
          (b) Optimized cost changes designed to halve the cumulative traffic count on the target link.
          (c) Cumulative traffic counts on the target link with and without control (cost changes).
          The dashed line represents the target value (50\% reduction).
   }
  \label{fig:chicago_control}
\end{figure}

The one-hour-ahead traffic nowcast predicted that the target link highlighted in Fig.~\ref{fig:chicago_control}a would experience the maximum cumulative traffic count of 3{,}030 among all links. 
We therefore considered a traffic control problem in which we seek a road pricing policy that reduces the future cumulative traffic count on this target link by half.
To this end, we defined a loss function as the squared error between the halved target count and the simulated count on the target link, and examined whether gradient-based optimization can solve this complex control problem, in which agents' stochastic route choices and their interactions are indirectly controlled through cost changes.

The resulting pricing pattern is not a simple uniform increase in costs around the target link, but rather a complex spatial structure that would be difficult to obtain through manual design (Fig.~\ref{fig:chicago_control}b).
While the optimization result shows intuitive trends, such as increasing the cost of the target link, it also assigns a non-trivial combination of cost increases and discounts to links around the target link.
The differentiable simulation techniques developed in this study enable gradients to propagate across links and automatically and efficiently adjust the costs of relevant links that contribute to high traffic counts on the target link, solely based on the desired traffic condition.
The optimized pricing policy achieved a reduction of 1{,}800 vehicles in the cumulative traffic count on the target link, corresponding to a 59\% decrease (Fig.~\ref{fig:chicago_control}c).
Because of the stochasticity of agents' behavior, there remains a gap of 9\% between the desired and achieved counts; nevertheless, the results indicate that the gradient signal effectively steers the traffic system toward the desired state.
Solving this control problem required only 728~s, demonstrating the efficacy of the differentiable model in addressing complex traffic control tasks via efficient gradient-based optimization.

\section*{Discussion}
We developed a suite of differentiable simulation techniques and, taken together, they constitute a differentiable agent-based traffic simulator that scales to one million vehicles on a city-scale network.
The combination of ultra-fast forward simulation and efficient gradient-based optimization enabled by our differentiable model allowed us to complete the full calibration--nowcast--control loop for one-hour-ahead traffic flows in about 20 minutes, leaving roughly 40 minutes of lead time for implementing interventions.
Conventional fine-grained simulation models are not differentiable and therefore must rely on inefficient gradient-free optimization methods for calibration and policy exploration~\cite{ma_2002,shultz_2004,park_2005,ma_2007,lee_2009}, which has hindered their use in real-time applications.
The dependence on gradient-free optimization has also constrained the scale and fidelity of computational models, because derivative-free methods typically require rapidly increasing numbers of function evaluations as the dimensionality and complexity of continuous optimization problems grow, making it difficult to obtain high-quality solutions within a realistic evaluation budget~\cite{rios_2013}.
A pioneering study proposed a computational model for a traffic digital twin of an expressway using real-time data~\cite{kusic_2023}.
However, the model is limited in scale and, critically, lacks a feedback loop from the computational model to the real traffic system.
By definition, a digital twin requires an automatic closed loop between a real-world system and its computational model, so that both the physical system and the model are continuously updated; otherwise, the computational model is merely a digital model or shadow~\cite{fuller_2020}. 
Establishing such a loop for a traffic digital twin requires solving complex optimization problems for calibration and traffic control, and solving these problems on high-fidelity models in real time has been computationally infeasible with conventional simulation models that rely on gradient-free optimization algorithms.
Our work provides a practical computational approach to address this long-standing challenge and to realize a traffic digital twin in which the closed loop between the real-world system and the computational model can be completed in real time.

The computational techniques we developed for differentiable agent-based traffic simulation are useful not only for traffic simulations but also for broader agent-based social simulations.
Differentiable agent-based simulations have thus far been applied to various applications; however, they have been limited to simplified models of society~\cite{andelfinger_2021,andelfinger_2023,chopra_2023,dyer_2023,querabofarull_2025}, and no existing study models large-scale spatial dynamics of agents while incorporating their stochastic decision-making.
Our differentiable computing techniques, such as the trajectory grafting technique developed for differentiable vehicle transfers between links, enable us to scale models to practical large-scale problems while maintaining end-to-end differentiability.
These techniques can be applied to build differentiable versions of a wide range of agent-based models, because decision making and spatial movement are typical modeling elements in agent-based simulations~\cite{bonabeau_2002}.
This would allow such models to exploit gradient information and powerful gradient-based optimization algorithms, ultimately facilitating the development of digital twins of societal systems.

With a broad range of extensions, the new computational foundation presented in this paper can serve as a core for solving real societal challenges.
A natural next step is to deploy the proposed framework continuously on real-world traffic networks and to assess its performance over repeated loops as it executes the cycle of calibration, nowcasting, and intervention.
For real deployments, challenges such as noisy and missing sensor data, feedback effects from behavioral changes induced by interventions, and the need for robust optimization that aligns with operational, safety, and equity constraints will arise.
As demonstrated on synthetic data, our model already has sufficient speed and scale for real-world applications; the remaining obstacles lie in addressing these deployment challenges, which constitute a distinct class of research questions at the interface of data assimilation, control theory, and human-in-the-loop system design.
To capture the full complexity of the real world, the model itself should be further enhanced.
Here, the differentiable nature of our framework opens up the possibility of joint optimization with other differentiable modules, such as demand forecasting~\cite{guo_2020} and signal control models~\cite{li_2016}, as the composition of differentiable models remains end-to-end differentiable and amenable to gradient-based optimization.
For example, jointly optimizing demand generation, mode choice, network flow, and signal policies under a single objective could blur the traditional separation between strategic transport planning and operational traffic control.
Beyond that, incorporating models of emissions, safety risk, or equity in a differentiable form would pave the way for multi-objective optimization directly targeting socially desirable outcomes.
When continuous assimilation of real-world data, fast nowcasting, and joint optimization of diverse subsystems are brought together, advanced real-time control of urban traffic networks becomes a technically tangible prospect.
The computational core of the traffic digital twin demonstrated in this work may thus serve as a foundation for future societal-scale digital twins, providing real-time decision-making infrastructure to address complex urban challenges, from congestion and emissions to resilience and accessibility.

\section*{Methods}
\subsection*{Newell's simplified car-following model}
Our model relies on Newell's simplified car-following model~\cite{newell_2002} to determine vehicle movements within links.
Specifically, we employed the following mesoscopic extension of the model~\cite{laval_2013,seo_2023,seo_2025}:
\begin{equation}
  \label{newell_model}
  x(t+{\Delta}t, n) = \min
  \left\{\!\begin{aligned}
&x(t, n) + u{\Delta}t \\[1ex]
&x(t+{\Delta}t-{\tau}{\Delta}n, n-{\Delta}n) - {1/{\kappa}}{\Delta}n\\[1ex]
\end{aligned}\right\} ,
\end{equation}
where $x(t, n)$ denotes the position of $n$-th platoon (agent) at time $t$, $u$ denotes the free-flow speed of the running link, $\tau$ denotes the reaction time, $\kappa$ denotes the jam density, ${\Delta}n$ denotes the platoon size, and ${\Delta}t$ denotes the simulation time step, which is given by $\tau{\Delta}n$.
In all experiments, we set $\tau = 1$.
This model determines a vehicle's position as the minimum of two potential positions: one derived from its free-flow state (the first term) and another from its congested state (the second term).
Conceptually, this formulation means that vehicles travel at their maximum free-flow velocity $u$, subject to the constraint of maintaining a safe distance from the vehicle ahead (defined by $1/{\kappa}{\Delta}n$).

\subsection*{Differentiable car-following model}
We have developed a fully tensorized algorithm for Newell's car-following model that leverages vectorization and masked tensor operations to update all agent positions in parallel (see Supplementary Information~A for implementation details).
Our formulation entirely avoids conditional statements and loops, rendering the algorithm highly amenable to JIT compilation.
This allows a JIT compiler to translate the operations into highly optimized kernels for modern hardware accelerators, unlocking significant performance gains.
Furthermore, the differentiability of every operation permits the use of efficient, gradient-based schemes via automatic differentiation, resulting in ultrafast optimization.
By design, our algorithm maintains numerical equivalence with the original model, thereby preserving fundamental traffic characteristics, such as ensuring agents do not move backward and remain strictly within link boundaries.
The model is implemented using JAX~\cite{frostig_2018, jax2018github}.

\subsection*{Differentiable node model}
Our node model governs agent navigation at intersections through a two-step discrete choice process, comprising an agent's link choice followed by a merge decision.
Crucially, this formulation is not only expressed in a fully tensorized form for simultaneous processing of all intersections, but is also fully differentiable despite the probabilistic and discrete nature of the choices.
Our choice model builds on standard discrete choice theory~\cite{train_2003}, and we develop a differentiable sampling scheme for this model using various differentiable computing techniques, such as the Gumbel–Softmax relaxation~\cite{jang_2017,maddison_2017} and the straight-through estimator~\cite{bengio_2013} (see Supplementary Information~B for implementation details).
This enables efficient gradient-based optimization of behavioral model parameters, even in the presence of agents' stochastic decision-making in the simulation.
Our implementation is built using JAX~\cite{frostig_2018, jax2018github}.

\subsection*{Differentiable state updates}
We developed a computational technique, which we term Trajectory Grafting (TG), to handle discontinuous updates of agent positions within a differentiable simulation.
In our simulation, an agent's position is updated discontinuously, for instance, when it moves from the end of one road link to the start of another.
If this transition is handled naively, the computational graph is broken at the point of the update.
This break disrupts the gradient flow required for automatic differentiation and prevents backpropagation across links.
The TG technique addresses this issue by maintaining a connected computational graph, which enables gradients to propagate through previous links and facilitates inter-link optimization (see Supplementary Information~C).

\subsection*{Optimization}
The same optimization algorithm is consistently used for both calibration and control problems.
It consists of four iterative steps repeated until convergence: (1) forward simulation, (2) loss calculation, (3) gradient calculation, and (4) parameter update.

In the forward simulation step, the differentiable traffic simulation is run to obtain traffic state estimates $\hat{\mathbf{y}}$.
Storing the entire computational graph for the subsequent automatic differentiation, however, can lead to prohibitive memory usage, especially for long-duration or large-scale simulations.
To address this, we employ checkpointing~\cite{margossian_2019}, a technique that reduces memory requirements for the backward pass by recomputing intermediate values instead of storing them.
Specifically, we insert a checkpoint at every time integration step using the {\texttt{jax.checkpoint}} function.
Furthermore, the time integration loop is implemented with {\texttt{jax.lax.scan}} to prevent loop unrolling and reduce JIT compilation time.
The loss function is the mean squared error between the estimated and observed traffic counts, where the squared errors are computed for each link with available observational data and then averaged over those links.
Because traffic counting is not naively differentiable, we develop a differentiable traffic counting technique for the loss calculation (see Supplementary Information~D).
The loss function, which includes a forward simulation to evaluate the gap, is JIT-compiled with {\texttt{jax.jit}} to enable efficient optimization.

The gradient of the loss function with respect to the parameters to be optimized is then computed.
As our simulator is end-to-end differentiable, this is achieved efficiently via automatic differentiation using the {\texttt{jax.grad}} function.
The parameters are subsequently updated at the end of each iteration using the computed gradients.
For this purpose, we employed the AdamW optimizer~\cite{loshchilov_2018} implemented using the Optax library~\cite{optax}, using default hyperparameters except for a learning rate of $10^{-1}$ and a weight decay of $10^{-5}$.
The optimization process is terminated when the loss has not improved in the most recent 20 iterations, and the parameters yielding the lowest loss are adopted as the final estimates.

\subsection*{Experimental setting}
We verified our model on the Sioux Falls network and the Chicago Sketch network, popular test problems in the transportation literature deposited in the Transportation Networks repository~\cite{tntp}.
To generate heterogeneous trip patterns, we constructed the networks by removing either the virtual inflow or outflow link connected to each node.
For terminal nodes that form the boundary of the simulation area (dead-end nodes), we did not remove the virtual links and retained both the inflow and outflow links to ensure that vehicles can exit the computational domain.
The resulting numbers of nodes/links for the Sioux Falls and Chicago networks are 48/100 and 993/2571, respectively (see Supplementary Information~F for a visual representation of the networks).

As each link has four parameters, i.e., a free-flow velocity $u$ (m/s), a jam density $\kappa$ (veh/m), a behavioral model parameter $\beta$, and a merge priority $\alpha$, the total numbers of parameters to be estimated for the Sioux Falls and Chicago networks are 400 and 10{,}284, respectively.
The cost parameter $c$ of each link was fixed uniformly at 1 during the calibration process, but was optimized only during the control process as a measure of intervention.

We evenly distributed 20{,}000 and 1{,}000{,}020 vehicles over the virtual inflow links of the Sioux Falls and Chicago networks, respectively, and simulated their dynamics for 90~min with randomized parameters to synthesize observation data.
These parameters used for synthesizing observations were randomly sampled from a uniform distribution over a realistic range, i.e., $u \in [13.9, 22.2]$, $\kappa \in [0.18, 0.22]$, $\beta \in [0, 5]$, and $\alpha \in [0.01, 5]$.
We simulated the traffic dynamics with ${\Delta}n = 1$ and ${\Delta}n = 30$ in the Sioux Falls and Chicago networks, respectively.

As observations for calibration, we assumed that cumulative traffic counts, measured at the midpoint of each link, are available every 5~min over the past 30~min.
To emulate realistic observation conditions, we added 10\% random noise to the traffic counts and assumed that these noisy observations are only partially available, i.e., on 80\% of the roads.
In the calibration process, the estimation starts from the mean values of the uniform parameter ranges, and the parameters are estimated from the observations using a gradient-based optimization scheme.
A forward simulation for the remaining 60~min is then conducted to evaluate performance in the traffic nowcasting and control problems.

All experiments were conducted on a computational server equipped with two Intel Xeon Platinum 8558 CPUs, utilizing a single NVIDIA H200 SXM 141GB GPU for acceleration.

\section*{Acknowledgements}
We thank the developers and contributors of the open-source simulator UX Sim~\cite{seo_2025}, which provided useful insights and inspiration during the development of our model.
We also acknowledge the Transportation Networks repository~\cite{tntp} for the network data used to construct the Sioux Falls and Chicago Sketch networks in this study.

\section*{Supplementary Information}

\setcounter{figure}{0}
\renewcommand{\thefigure}{S\arabic{figure}}

\setcounter{algorithm}{0}
\renewcommand{\thealgorithm}{S\arabic{algorithm}}

\subsection*{A. Differentiable car-following model}
Throughout our algorithm, agent positions are represented as the following two-dimensional tensor $\mathbf{X} \in \mathbb{R}^{|\mathcal{N}| \times |\mathcal{L}|}$:
\begin{equation}
  X_{ij} =
  \begin{cases}
    x_{ij} & \text{if agent $i$ is on link $j$}, \\
    -M & \text{otherwise},
  \end{cases}
\end{equation}
where $\mathcal{N}$ and $\mathcal{L}$ denote the sets of agents and links, respectively, $x_{ij}$ represents the position of the agent $i$ on the link $j$, and $M$ is a large constant (e.g., $99999$).
Updating all vehicles simultaneously by processing the large tensor $\mathbf{X}$ is memory-demanding, particularly in large-scale applications.
Therefore, we adopted a per-link computational approach by vectorizing the car-following model using the {\texttt{vmap}} function in JAX.
This strategy enables parallel vehicle updates while substantially reducing memory usage.
The following describes this per-link computation, which operates on $\mathbf{x}^{(j)}$, a $j$-th column of the tensor $\mathbf{X}$.

We first calculate the position updates for the free-flow state, ${\Delta}\mathbf{x}^{(j)}_{free}$, as follows:
\begin{equation}
  \label{first_term}
  {\Delta}\mathbf{x}^{(j)}_{free} = \mathbb{I}_{\mathbf{x}^{(j)}\ge0} \circ \mathbf{u}^{(j)} \circ {\Delta}t, 
\end{equation}
where $\mathbf{u}^{(j)} \in \mathbb{R}^{|\mathcal{N}|}_{\ge 0}$ represents the free flow speed within the link $j$, $\mathbb{I}_{\mathbf{x}^{(j)}\ge0} \in \{0, 1\}^{|\mathcal{N}|}$ is an indicator tensor which masks $-M$ in $\mathbf{x}^{(j)}$, $\circ$ denotes the Hadamard product.
In our model, each link $j$ has a free-flow velocity $u_j$.
Agents traveling on a given link adopt its corresponding velocity.
By applying these masked operations, we can simultaneously compute the free-flow position updates for all vehicles.

Next, we calculate the position updates for vehicles under congested conditions.
The next position of vehicle $n$ is determined relative to the preceding vehicle, $x(t, n-1)$.
By leveraging the identity $x(t, n-1) = x(t, n) + h(t, n)$, where $h(t, n)$ is the space headway, we can express the next position in terms of vehicle $n$'s own state.
This yields the expression $x(t, n) + h(t, n) - 1/{\kappa}{\Delta}n$.
Therefore, once the space headway of each vehicle is determined, we can compute the position updates for the congested state, ${\Delta}\mathbf{x}^{(j)}_{cong}$.
Rather than relying on the computationally expensive, brute-force method of checking all pairwise vehicle distances, we constructed the space headway tensor $\mathbf{h}^{(j)} \in \mathbb{R}^{|\mathcal{N}|}$ efficiently by sorting the vehicles according to their position.
To obtain the sorted indices of valid vehicles on the link $j$, we first compute an auxiliary tensor $\tilde{\mathbf{x}}$ as follows:
\begin{equation}
    \tilde{\mathbf{x}} = \mathbb{I}_{\mathbf{x}^{(j)} \ge 0} \circ \mathbf{x}^{(j)} - \mathbb{I}_{\mathbf{x}^{(j)} < 0} \circ M.
\end{equation}
We then obtain the sorted indices $\boldsymbol{\sigma}$ by
\begin{equation}
  \boldsymbol{\sigma} = \mathrm{argsort}(-\tilde{\mathbf{x}}).
\end{equation}
The indexing operation, using the output of the {\texttt{argsort}} function, does not disrupt the computational graph, preserving the path for gradient backpropagation.
The sorted indices $\boldsymbol{\sigma}$ are then used to compute the components $h'_{i}$ of the preliminary space headway tensor $\mathbf{h}'$ as follows:
\begin{equation}
   h'_i =
  \begin{cases}
    x^{(j)}_{\sigma_{k-1}} - x^{(j)}_{\sigma_{k}} & \text{if $i=\sigma_k$ for some $k \in \{2,\ldots,|\mathcal{N}|\}$}, \\
    M & \text{otherwise}.
  \end{cases}
\end{equation}
While this procedure yields correct headways for valid vehicles, it produces spurious values for invalid ones.
This is because the placeholder value $M$ assigned to their positions renders the subtraction-based calculation meaningless.
Therefore, we obtain the final space headway tensor $\mathbf{h}^{(j)}$ by correcting these spurious values in $\mathbf{h}'$ as follows:
\begin{equation}
  \mathbf{h}^{(j)} = \mathbb{I}_{\mathbf{x}^{(j) \ge 0}} \circ \mathbf{h}' + \mathbb{I}_{\mathbf{x}^{(j)} < 0} \circ M.
\end{equation}
This artificially large headway is designed to manipulate the state-selection mechanism.
It renders the congested-state update, ${\Delta}\mathbf{x}^{(j)}_{cong}$, extremely large for the leading vehicle.
Since the final position update is determined by $\min({\Delta}\mathbf{x}^{(j)}_{free}, {\Delta}\mathbf{x}^{(j)}_{cong})$, this forces the model to select the free-flow update, ${\Delta}\mathbf{x}^{(j)}_{free}$, for leading vehicles.
In contrast, all other following vehicles operate with finite, realistic headways, allowing their states to be determined by the actual local traffic conditions.
In the congested state, a small space headway $\mathbf{h}^{(j)}$ can cause the position update ${\Delta}\mathbf{x}^{(j)}_{cong}$ to become negative, implying an unphysical backward movement.
To preclude this possibility and enforce non-negative updates, we compute ${\Delta}\mathbf{x}^{(j)}_{cong}$ as follows:
\begin{equation}
  {\Delta}{\mathbf{x}^{(j)}_{cong}} = \mathbb{I}_{\mathbf{x}^{(j)}\ge0} \circ \max(\mathbf{h}^{(j)} - 1/{\boldsymbol{\kappa}^{(j)}}{\Delta}n, \mathbf{0}).
\end{equation}
Here, $\boldsymbol{\kappa}^{(j)} \in \mathbb{R}^{|\mathcal{N}|}$ denotes the tensor of jam densities on link $j$, $\mathbf{0} \in \{0\}^{|\mathcal{N}|}$ is the zero tensor.
The indicator function $\mathbb{I}_{\mathbf{x}^{(j)}\ge0}$ serves as a mask to nullify the updates for invalid vehicles.

The position update ${\Delta}\mathbf{x}^{(j)}$ is then calculated as the element-wise minimum of the free-flow update ${\Delta}\mathbf{x}^{(j)}_{free}$ and the congested-flow update ${\Delta}\mathbf{x}^{(j)}_{cong}$:
\begin{equation}
  {\Delta}\mathbf{x}^{(j)} = \min({\Delta}\mathbf{x}^{(j)}_{free}, {\Delta}\mathbf{x}^{(j)}_{cong}). 
\end{equation}
Subsequently, the preliminary position at the next time step, ${\mathbf{x}^{(j)}}^{\prime}(t + \Delta t)$, is determined as:
\begin{equation} 
  {\mathbf{x}^{(j)}}^{\prime}(t + \Delta t) = \mathbf{x}^{(j)}(t) + {{\Delta}{\mathbf{x}}}^{(j)}.
\end{equation}
If the displacement $\Delta{\mathbf{x}^{(j)}}$ is large, this preliminary position may exceed the length of the current link.
To ensure that the final position $\mathbf{x}^{(j)}(t + \Delta t)$ remains within the link's boundaries, a correction is necessary.
This correction utilizes a position-limiting tensor, $\mathbf{x}^{(j)}_{limit}$, constructed as follows: 
\begin{equation}
  \label{eq:tg}
  \mathbf{x}^{(j)}_{limit} = \mathbf{L}^{(j)} + {\mathbf{x}^{(j)}}^{\prime} - \mathrm{stop\_gradient}({\mathbf{x}^{(j)}}^{\prime}),
\end{equation}
where $\mathbf{L}^{(j)} \in \mathbb{R}^{|\mathcal{N}|}$ is a constant tensor with all element equal to the length of link $j$.
The $\mathrm{stop\_gradient}(\cdot)$ operator blocks the gradient from flowing through its argument during backpropagation, effectively treating the argument as a constant.
A naive clipping of the position, for instance, by simply replacing ${\mathbf{x}^{(j)}}^{\prime}$ with $\mathbf{L}^{(j)}$, would detach the operation from the computational graph.
This is because $\mathbf{L}^{(j)}$ is a constant and carries no gradient information, which would prevent gradients from being propagated back to earlier computations.
To address this, we introduce a technique inspired by the straight-through estimator~\cite{bengio_2013}, which we term Trajectory Grafting (TG).
In Eq.~\ref{eq:tg}, the term ${\mathbf{x}^{(j)}}^{\prime} - \mathrm{stop\_gradient}({\mathbf{x}^{(j)}}^{\prime})$ serves to isolate the computational trajectory of ${\mathbf{x}^{(j)}}^{\prime}$, effectively acting as a gradient carrier.
By adding this gradient carrier to the constant link length, we graft the computational history of ${\mathbf{x}^{(j)}}^{\prime}$ onto the new limiting value.
Therefore, the TG technique allows us to cap the agent positions at the link boundaries without breaking the computational graph essential for gradient-based optimization.
Using this limiting tensor, the final position at the next time step is determined by taking the element-wise minimum with the preliminary position:
\begin{equation}
  \mathbf{x}^{(j)}(t + \Delta t) = \min({\mathbf{x}^{(j)}}^{\prime}, \mathbf{x}^{(j)}_{limit}).
\end{equation}
By mapping each newly computed tensor $\mathbf{x}^{(j)}(t + \Delta t)$ back to the $j$-th column of the original tensor $\mathbf{X}$, we obtain the updated tensor $\mathbf{X}(t + \Delta t)$.
We provide the pseudocode of the differentiable car-following model in Alg.~\ref{alg:carfollowingmodel}.

\begin{algorithm}[H]
\caption{Per-link differentiable car-following model.}
\label{alg:carfollowingmodel}
\begin{lstlisting}
def compute_car_following_model(x, u, kappa, L, delta_t, delta_n, M):

  # mask for invalid vehicles
  I_xpos = x >= -1e-2

  # obtain dx in free-flow state
  dx_free = (u * delta_t) * I_xpos

  # obtain dx in congested state
  x_tilde = where(I_xpos, x, -1e12)
  sigma = argsort(-x_tilde)
  front = array([M], dtype=x_tilde.dtype)
  prev = x_tilde[:-1]
  curr = x_tilde[1:]
  diffs = prev - curr
  diffs_full = concatenate([front, diffs])
  headway = full_like(x, M)
  headway = headway.at[sigma].set(diffs_full)
  headway = where(I_xpos, headway, M)
  dx_cong = maximum(headway - 1.0 / kappa * delta_n, 0) * I_xpos

  # determine dx
  dx = minimum(dx_free, dx_cong)

  # update x
  x_new = x + dx

  # limit x within link
  x_limit = L + x_new - stop_gradient(x_new)  
  x_new = minimum(x_new, x_limit)

  return x_new

\end{lstlisting}
\end{algorithm}

\subsection*{B. Differentiable node model}
\subsubsection*{Link choice}
First, the node model determines the agents' link choice.
This process begins by identifying the set of potential next links for each agent.
This set, $\mathcal{L}^{*} \in \{0, 1\}^{|\mathcal{N}| \times |\mathcal{L}|}$, is expressed as:
\begin{equation}
  \mathcal{L}^{*} = \mathbb{I}_{\mathbf{X}\ge0} \cdot \mathbf{A},
\end{equation}
where $\mathbf{A} \in \{0, 1\}^{|\mathcal{L}| \times |\mathcal{L}|}$ is the network's adjacency tensor, where the element $A_{ij} = 1$ if link $j$ is reachable from link $i$, and $0$ otherwise.
$\mathbf{A}$ is static and determined by the road network geometry.
Using $\mathcal{L}^{*}$, we compute the link utility tensor $\mathbf{V} \in \mathbb{R}^{|\mathcal{N}| \times |\mathcal{L}|}$ as:
\begin{equation}
  \mathbf{V} = (\boldsymbol{\beta} \circ 1/\mathbf{c}) * \mathcal{L}^{*} - M \circ (\bar{\mathcal{L}^{*}}),
\end{equation}
where $\bar{\mathcal{L}^{*}}$ represents the complement of $\mathcal{L}^{*}$, $\boldsymbol{\beta} \in \mathbb{R}^{|\mathcal{L}|}_{\ge 0}$ is the learnable behavioral parameter that determines how the travel cost $\mathbf{c} \in \mathbb{R}^{|\mathcal{L}|}_{> 0}$ affects link choice, and the operator $*$ denotes an element-wise product with broadcasting.
Consequently, the utility $V_{ij}$ for agent $i$ choosing the link $j$ becomes $\beta_j / c_j$ if the transition is possible and a large negative number $-M$ otherwise, effectively restricting the choice to valid links.
While we present a simple utility function based on the cost factor with $\boldsymbol{\beta}$, our formulation is general enough to incorporate additional factors, such as those related to an agent's origin and destination.
From the utility tensor $\mathbf{V}$, the choice probability for each agent over all possible next links, $\mathbf{z} \in [0, 1]^{|\mathcal{N}| \times |\mathcal{L}|}$, is computed using a softmax function applied along the link dimension (the second axis):
\begin{equation}
  \label{eq:softmax}
  z_{ij} = \frac{\mathrm{exp}(V_{ij})}{\sum_{k=1}^{|\mathcal{L}|}\mathrm{exp}(V_{ik})}.
\end{equation}

To obtain a discrete link choice for each agent, one might sample from the probability distribution $\mathbf{z}$, for instance by taking the {\texttt{argmax}}.
However, such a sampling process is non-differentiable, which obstructs the end-to-end gradient flow essential for automatic differentiation.
To address this challenge, we employ a two-stage approach.
First, we use the Gumbel-Softmax trick~\cite{jang_2017,maddison_2017} to generate a differentiable proxy for discrete samples.
This yields the following soft choice tensor $\boldsymbol{\pi} \in [0, 1]^{|\mathcal{N}| \times |\mathcal{L}|}$:
\begin{equation}
  \label{eq:gumbel}
  \pi_{ij} = \frac{\mathrm{exp}((\mathrm{log}(z_{ij})+g_{ij})/{\tau})}{\sum_{k=1}^{|\mathcal{L}|} \mathrm{exp}((\mathrm{log}(z_{ik}) + g_{ik})/{\tau})},
\end{equation}
where $g_{ij}$ are i.i.d samples drawn from a $\mathrm{Gumbel}(0,1)$ distribution, and $\tau$ is a temperature parameter, which we set to $0.01$.
While $\boldsymbol{\pi}$ is differentiable, it consists of continuous values, whereas our simulation requires discrete, one-hot choices.
To bridge this gap, we next apply the straight-through estimator~\cite{bengio_2013}, allowing us to use discrete samples in the forward pass while preserving a valid gradient in the backward pass.
The preliminary link choice tensor $\mathbf{l}' \in \{0, 1\}^{|\mathcal{N}| \times |\mathcal{L}|}$ is then obtained as follows:
\begin{equation}
  \label{eq:straigththrough}
  \mathbf{l}' = \mathrm{onehot}(\boldsymbol{\pi}) + \boldsymbol{\pi} - \mathrm{stop\_gradient}(\boldsymbol{\pi}),
\end{equation}
where $\mathrm{onehot}$ operator converts the input into a one-hot tensor by applying {\texttt{argmax}} to the soft link-choice samples along the second axis.
These techniques enable us to use exact discrete sampling in forward simulation while its gradient is seamlessly passed through the continuous proxy $\boldsymbol{\pi}$ in the backward pass.

A sampled link choice is considered valid only if the following three conditions are met: the current link is not a dead-end, the agent has reached the end of its current link, and the chosen destination link has sufficient capacity.
To enforce this, we correct $\mathbf{l}'$ to obtain the final, valid link choice tensor $\mathbf{l}$ by applying three indicator masks: a connectivity mask $\mathbb{I}_{connected} \in \{0, 1\}^{|\mathcal{L}|}$, an arrival mask $\mathbb{I}_{arrived} \in \{0, 1\}^{|\mathcal{N}|}$, and a vacancy mask $\mathbb{I}_{vacant} \in \{0, 1\}^{|\mathcal{L}|}$. 
Using these masks, the final link choice tensor $\mathbf{l} \in \{0, 1\}^{|\mathcal{N}| \times |\mathcal{L}|}$ is computed as:
\begin{equation}
  \mathbf{l} = \mathbb{I}_{connected} * (\mathbb{I}_{arrived} * (\mathbb{I}_{vacant} * \mathbf{l}')).
\end{equation}
The connectivity mask, $\mathbb{I}_{connected}$, is obtained by applying a {\texttt{max}} function to $\mathcal{L}^{*}$ along its second axis.
The arrival mask, $\mathbb{I}_{arrived}$, is computed by applying a {\texttt{sum}} function to $\mathbb{I}_{\mathbf{X}=\mathbf{L}}$ along the second axis, where $\mathbf{L} \in \mathbb{R}^{|\mathcal{N}| \times |\mathcal{L}|}$ is the link length tensor.
A link is deemed vacant if the rearmost agent on it is sufficiently far from the entry point.
The position of this rearmost agent, $\mathbf{X}_{min} \in \mathbb{R}^{|\mathcal{L}|}$, is found by applying a {\texttt{min}} function to the masked tensor $\mathbb{I}_{\mathbf{X} \ge 0} {\circ} \mathbf{X} + \mathbb{I}_{\mathbf{X} < 0}{\circ}M$ along the first axis. 
The vacancy mask is then given by $\mathbb{I}_{vacant} = \mathbf{X}_{min} > 1/{\boldsymbol{\kappa}}{\Delta}n$.

\subsubsection*{Merge choice}
Next, we address the merge choice problem, which resolves merging conflicts when multiple agents, based on the link choice result $\mathbf{l}$, compete to enter the same downstream link.
To this end, we introduce a merge priority $\boldsymbol{\alpha} \in \mathbb{R}^{|\mathcal{L}|}_{> 0}$ that represents the intrinsic priority for merging from each link.
The merging priority tensor $\mathbf{p} \in \mathbb{R}^{|\mathcal{L}| \times |\mathcal{N}|}$ is then formulated as:
\begin{equation}
  \mathbf{p} = ((\mathbb{I}_{\mathbf{X} \ge 0} \cdot \boldsymbol{\alpha}) * \mathbf{l})^{\mathsf{T}}.
\end{equation}
In a similar manner to the construction of the link utility tensor $\mathbf{V}$, we construct a merging utility tensor $\mathbf{V}' \in \mathbb{R}^{|\mathcal{L}| \times |\mathcal{N}|}$ for this second problem as follows:
\begin{equation}
  \mathbf{V}' = \mathbf{p} - \mathbb{I}_{\mathbf{p}=0} \circ M.
\end{equation}
An entry $V'_{ij}$ becomes the merging priority of agent $j$ for link $i$ and $-M$ otherwise.
By applying the same sampling procedures (Eqs.~\ref{eq:softmax}-\ref{eq:straigththrough}) to $\mathbf{V}'$, we obtain a preliminary sampling of transferring agents, $\mathbf{a}' \in \{0, 1\}^{|\mathcal{L}| \times |\mathcal{N}|}$, where $a'_{ij} = 1$ if agent $j$ is chosen to transfer to link $i$, and 0 otherwise.
However, invalid sampling occurs when no agent targets a particular link $i$, as the corresponding utilities $V'_{ij}$ for all agents $j$ become uniformly $-M$.
To prevent this, we mask the preliminary sampling result $\mathbf{a}'$ to yield the valid sampling of transferring agents $\mathbf{a} \in \{0, 1\}^{|\mathcal{L}| \times |\mathcal{N}|}$:
\begin{equation}
  \mathbf{a} = (\max(\mathbf{l}))^{\mathsf{T}} * \mathbf{a}',
\end{equation}
where the {\texttt{max}} function is applied along the first axis.
By solving these successive two-step choice problems (link and merge choice), we can correctly simulate probabilistic yet consistent inter-link vehicle transfers, while fully preserving end-to-end differentiability for gradient-based optimization.

\subsubsection*{Differentiable vehicle transfer}
To implement the vehicle transfers based on the choice result $\mathbf{a}$, we update the agent state tensor $\mathbf{X}$ in a differentiable manner.
This requires two auxiliary indicator tensors: $\mathbb{I}_{remove} \in \{0, 1\}^{|\mathcal{N}| \times |\mathcal{L}|}$ for removing agents from their original links, and $\mathbb{I}_{add} \in \{0, 1\}^{|\mathcal{N}| \times |\mathcal{L}|}$ for adding them to their new links.
The tensor $\mathbb{I}_{remove}$ identifies agents that are leaving their current link and is obtained by:
\begin{equation}
  \mathbb{I}_{remove} = \mathbb{I}_{\mathbf{X} \ge 0} * (\mathrm{sum}({\mathbf{a}}^{\mathsf{T}}))^{\mathsf{T}},
\end{equation}
where the {\texttt{sum}} function is performed along the second axis.
Similarly, $\mathbb{I}_{add}$ is given by the transpose of the choice result, $\mathbf{a}^{\mathsf{T}}$.

To reflect these inter-link transfers, we update $\mathbf{X}$.
A naive update would break the computational graph; therefore, we again employ the TG technique.
The inter-link transfers are implemented as follows:
\begin{equation}
  \label{eq:transfer}
  \mathbf{X} \leftarrow \mathbf{X} - (M + \mathbf{x}) * \mathbb{I}_{remove} + (M + \mathbf{x} - \mathrm{stop\_gradient}(\mathbf{x})) * \mathbb{I}_{add},
\end{equation}
where $\mathbf{x} \in \mathbb{R}^{|\mathcal{N}|}$ represents the tensor of valid vehicle positions, obtained by applying the {\texttt{max}} function along the first axis of $\mathbf{X}^{\mathsf{T}}$.
This operation updates an agent's position on its original link to $-M$ and initializes its position on the new link to 0.
The TG technique, represented by the term $\mathbf{x} - \mathrm{stop\_gradient}(\mathbf{x})$ in Eq.~\ref{eq:transfer}, grafts the computational trajectories on the previous link onto the new position.
If this update were performed without TG, the computational history would be lost, preventing backpropagation across previous links.
We provide the pseudocode of the differentiable node model in Alg.~\ref{alg:nodemodel}.

\begin{algorithm}[H]
\caption{Differentiable node model.}
\label{alg:nodemodel}
\begin{lstlisting}
def compute_node_model(x, alpha, beta, cost, kappa, L, A, M):

  # mask for invalid vehicles
  I_xpos = x >= -1e-2

  # potential next link
  L_star = matmul(I_xpos, A)
  L_star_bar = L_star < 1

  # compute utility for link choice sampling
  V = (beta * cost) * L_star - M * L_star_bar

  # link choice sampling
  z = softmax(V, axis=1)
  l_prime = gumbel_softmax_sample(z)

  # generate masks to correct sampling
  is_arrived = x >= L - 1e-2
  is_vacant = min(M * (x < -1e-2) + (x >= -1e-2) * x, axis=0)
  I_arrived = asarray([sum(is_arrived, axis=1)]).T
  I_vacant = is_vacant > 1 / kappa * delta_n
  I_connected = asarray([sum(L_star, axis=1)]).T

  # correct link choice sampling
  l = I_connected * (I_arrived * (I_vacant * l_prime))

  # compute priority
  p = (matmul(I_xpos, alpha) * l).T

  # compute utility for merge choice sampling
  I_pzero = p == 0
  V_prime = p - M * I_pzero

  # merge choice sampling
  z_prime = softmax(V_prime, axis=1)
  a_prime = gumbel_softmax_sample(z_prime)

  # correct merge choice sampling
  a = asarray([max(l, axis=0)]).T * a_prime
  
  # update x
  I_remove = I_xpos * asarray([sum(a.T, axis=1)]).T
  I_add = a.T
  x_prime = asarray([max(x, axis=1)]).T
  x_new = x - (M + x_prime) * I_remove + (M + x_prime - stop_gradient(x_prime)) * I_add

  return x_new

\end{lstlisting}
\end{algorithm}

\clearpage

\subsection*{C. Trajectory grafting}
We developed a computational technique, termed Trajectory Grafting (TG), to enable discontinuous state updates of agents while maintaining differentiability.
Such discontinuous updates frequently occur in traffic simulations, for example when overwriting vehicle positions with the link length to keep them within a link, or when resetting the position to zero on the downstream link after an inter-link transfer.
The key idea of TG is to graft the computational history accumulated in the old value $v_{old}$ onto a discontinuously updated new value $v_{new}$, so that gradients can still flow through the previous computational history.
This allows parameters involved in earlier computations to be effectively optimized via automatic differentiation, even in the presence of discontinuous updates.
TG can be implemented as
\begin{equation}
  v_{new} \leftarrow v_{new} + v_{old} - \mathrm{stop\_gradient}(v_{old}).
\end{equation}
The term $v_{old} - \mathrm{stop\_gradient}(v_{old})$ preserves the computaional history of the previous value and grafts it onto $v_{new}$, while not altering the numerical value of $v_{new}$.
In JAX, the $\mathrm{stop\_gradient}(\cdot)$ operator can be implemented using the {\texttt{jax.lax.stop\_gradient}} function.

\begin{figure}[H]
  \centering
  \includegraphics[width=\textwidth]{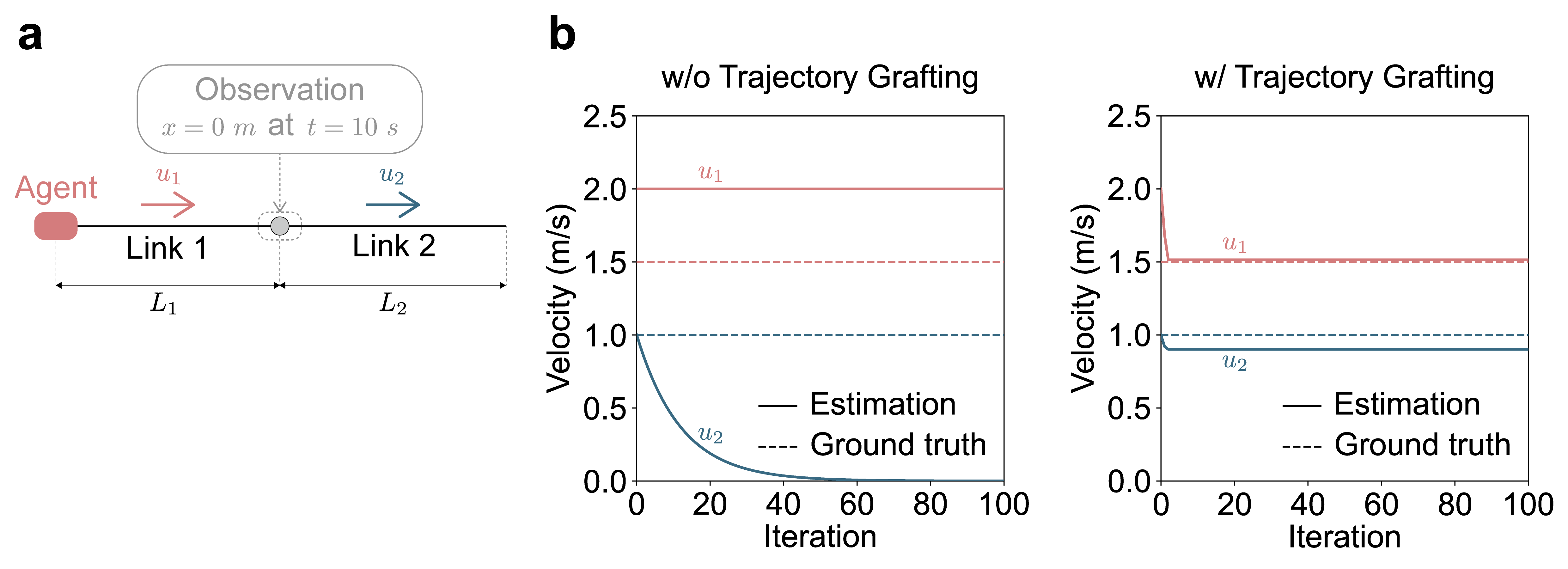}
  \caption{Effect of the trajectory grafting on parameter estimation.
           (a) Experimental setup.
           A single agent travels from Link 1 to Link 2 over a total duration of $t=10~\mathrm{s}$.
           The link lengths are $L_1 = L_2 = 15~\mathrm{m}$.
           The ground-truth trajectory is generated with free-flow speeds $u_1 = 1.5~\mathrm{m/s}$ on Link 1 and $u_2 = 1.0~\mathrm{m/s}$ on Link 2, resulting in a single observation of the agent at position $x=0~\mathrm{m}$ on Link 2 at $t=10~\mathrm{s}$.
           From this observation, we estimate $u_1$ and $u_2$ by minimizing the squared error between the observed and simulated positions.
           (b) Optimization results without TG (left) and with TG (right).
           The solid and dashed lines represent estimated and true values, respectively.
           The results for $u_1$ and $u_2$ are shown in different colors.
           }
  \label{fig:tg_experiment}
\end{figure}

In a simple toy problem, we verify the effect of the TG technique on optimization results.
Fig.~\ref{fig:tg_experiment}a illustrates the experimental setting.
In this problem, a single agent travels from Link 1 to Link 2 over a total duration of $10~\mathrm{s}$.
The agent moves with a free-flow speed of $u_1$ on Link 1 and $u_2$ on Link 2.
The lengths of Link 1 and Link 2 are $L_1 = 15~\mathrm{m}$ and $L_2 = 15~\mathrm{m}$, respectively.
For the ground truth, we synthesized the agent's trajectory with $u_1 = 1.5~\mathrm{m/s}$ and $u_2 = 1.0~\mathrm{m/s}$.
This setting yields a single observation of the agent at position $x=0~\mathrm{m}$ on Link 2 at time $t=10~\mathrm{s}$.
From this observation, we estimate both $u_1$ and $u_2$ so as to reproduce the observed state.
The initial values for the estimation were set to $u_1 = 2.0~\mathrm{m/s}$ and $u_2 = 1.0~\mathrm{m/s}$.
As a loss function $\ell$, we use the squared error between the observed and estimated positions.
The optimization was performed using a simple gradient descent algorithm with a fixed learning rate of $10^{-2}$, based on the gradients ${\partial}{\ell}/{\partial}{u_1}$ and ${\partial}\ell/{\partial}{u_2}$.

Fig.~\ref{fig:tg_experiment}b reports the optimization results.
Without TG, although the loss almost reaches zero, the optimization yields incorrect speed estimates of $u_1 \approx 2.0~\mathrm{m/s}$ and $u_2 \approx 0~\mathrm{m/s}$; the estimated system reproduces the observation by freezing the agent on Link 2.
This occurs because, in a naive state-update scheme, the computational history on Link 1 is discarded at the point of the link transfer, and the position of the agent on Link 2 depends only on $u_2$, i.e., ${\partial}{\ell}/{\partial}{u_1} = 0$.
In contrast, the model with TG successfully estimated the speeds as $u_1 \approx 1.5~\mathrm{m/s}$ and $u_2 \approx 0.9~\mathrm{m/s}$, which are close to the true values.
The TG technique grafts the computational history on Link 1 onto the state on Link 2, thereby correctly attributing a large contribution of $u_1$ to the position on Link 2 and guiding the optimization toward the true parameters.

\subsection*{D. Differentiable traffic counting}
We develop a differentiable traffic counting technique because counting is not naively differentiable.
Here, we describe the one-dimensional (per-link) case of the proposed approach for illustration.

Given the positions of agents on link $j$, $\mathbf{x}^{(j)} \in \mathbb{R}^{|\mathcal{N}|}$, and the location of the traffic counter $\mathbf{o}^{(j)} \in \mathbb{R}^{|\mathcal{N}|}$, we first compute the non-differentiable traffic count indicator $\mathbb{I}^{(j)}_{Q} \in \{0, 1\}^{|\mathcal{N}|}$, which indicates whether each vehicle has passed the observation point, as $\mathbb{I}^{(j)}_{Q} = \mathbf{x}^{(j)} \ge \mathbf{o}^{(j)}$.
We then compute the differentiable traffic count indicator $\mathbb{I}'_{Q} \in \mathbb{R}^{|\mathcal{N}|}$ as
\begin{equation}
  \mathbb{I}^{(j)'}_{Q} = \mathrm{sigmoid}(s(\mathbf{x}^{(j)} - \mathbf{o}^{(j)})),
\end{equation}
where $\mathrm{sigmoid}(\cdot)$ is the sigmoid function, and $s$ is a scale parameter.
In this study, we systematically determine the scale parameter using the length of link $j$, $L^{(j)}$, as $s = 5 / L^{(j)}$.

Using the non-differentiable indicator $\mathbb{I}^{(j)}_{Q}$ and the differentiable indicator $\mathbb{I}^{(j)'}_{Q}$, we construct a differentiable yet exact count indicator as $\mathbb{I}^{(j)}_{Q} + \mathbb{I}^{(j)'}_{Q} - \mathrm{stop\_gradient}(\mathbb{I}^{(j)'}_{Q})$.
Applying the sum function to this indicator yields the total count $Q^{(j)}$ on link $j$, i.e., $Q^{(j)} = \mathrm{sum}(\mathbb{I}^{(j)}_{Q} + \mathbb{I}^{(j)'}_{Q} - \mathrm{stop\_gradient}(\mathbb{I}^{(j)'}_{Q}))$.

\subsection*{E. Application to Sioux Falls network}
We applied the developed algorithm to traffic flows on the Sioux Falls network, synthesized with a platoon size of $\Delta n = 1$.
The same platoon size of $\Delta n = 1$ was consistently used in the simulations for estimating the ground-truth traffic flows.

\begin{figure}[H]
  \centering
  \includegraphics[width=\textwidth]{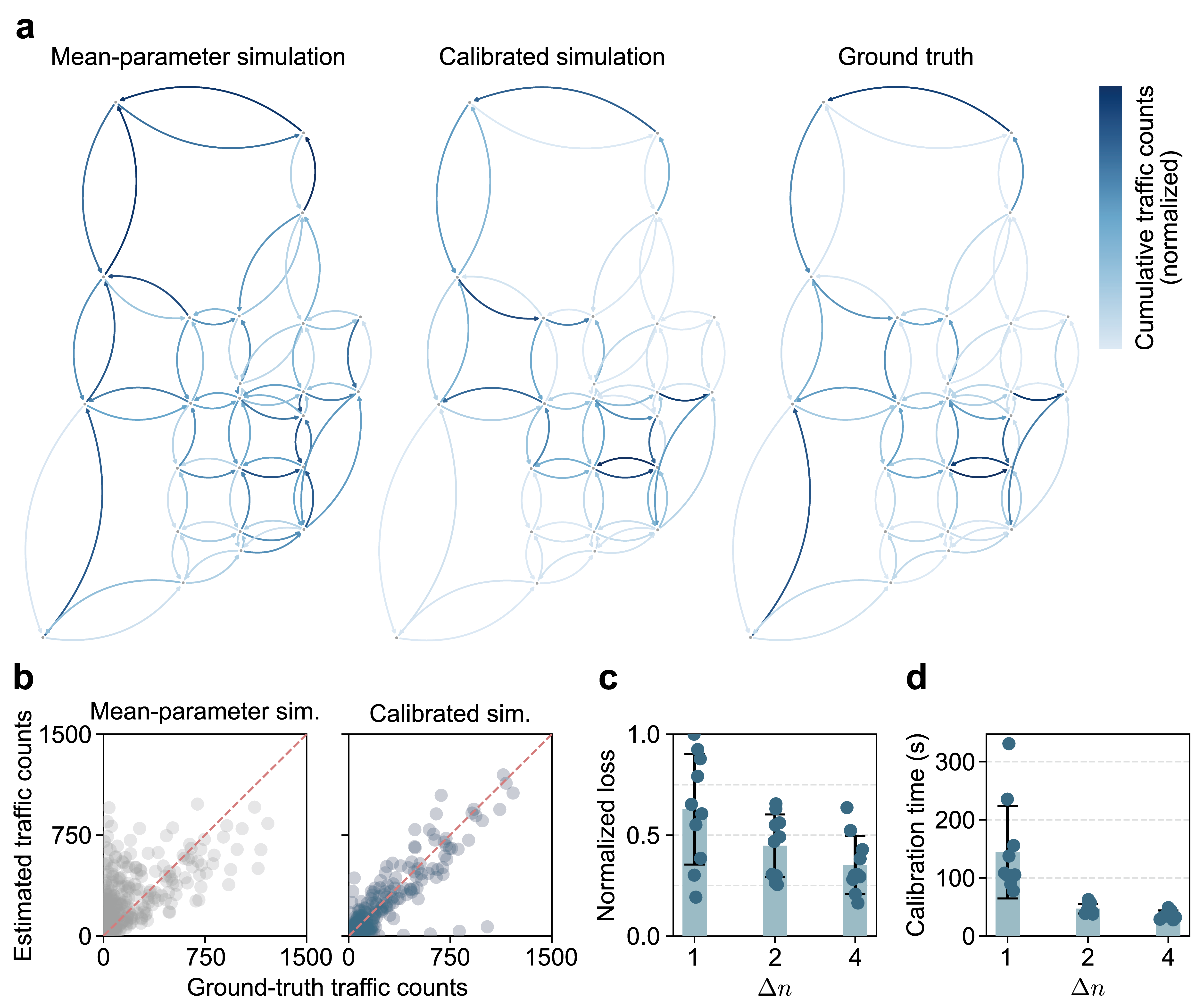}
  \caption{Calibration results on the Sioux Falls network.
          (a) Cumulative traffic counts from the mean-parameter simulation and the calibrated simulation compared against the ground truth.
          (b) Traffic counts (every 5 minutes) from the mean-parameter simulation and the calibrated simulation compared against the ground truth.
          (c) Effect of the platoon size $\Delta n$ on the loss (estimation quality). 
          (d) Effect of the platoon size $\Delta n$ on the calibration time. 
          The dots and whiskers represent the measured values and their standard deviations over 10 trials, respectively.
          }
          \label{fig:res_siouxfalls}
\end{figure}

The simulation with the gradient-based calibration successfully reproduced the spatially heterogeneous ground-truth traffic counts, whereas the mean-parameter simulation failed to reproduce the ground-truth traffic counts (Fig.~\ref{fig:res_siouxfalls}a).
The calibrated simulation quantitatively reproduced the ground-truth traffic flows (MAE of 47.5; Pearson's $r = 0.90$), whereas the mean-parameter simulation failed to capture the ground-truth traffic flows (MAE of 116.2; Pearson's $r = 0.66$), corresponding to a 59\% improvement in the traffic count estimation (Fig.~\ref{fig:res_siouxfalls}b).
This calibration using 30~min of traffic observations required 144.3~s on average over 10 trials, which corresponds to 12.5$\times$ real-time.

We further evaluated the performance of the gradient-based calibration with different platoon sizes $\Delta n$ and found that the calibration time can be dramatically reduced without degrading the loss (Fig.~\ref{fig:res_siouxfalls}c, d).
With larger $\Delta n = 2$ and $\Delta n = 4$, the calibration time was reduced to 47.0~s (3.1$\times$ speed-up) and 36.8~s (3.9$\times$ speed-up) on average over 10 trials, while the loss values were improved by 29\% and 44\%, respectively.
For larger $\Delta n$, we can take a larger time step $\Delta t$, because $\Delta t$ is determined by $\tau \Delta n$ with $\tau = 1$.
This leads to a shorter time-integration loop for simulating the target traffic flows and substantially accelerates the optimization process.

\subsection*{F. Road networks}
\begin{figure}[H]
  \centering
  \includegraphics[width=\textwidth]{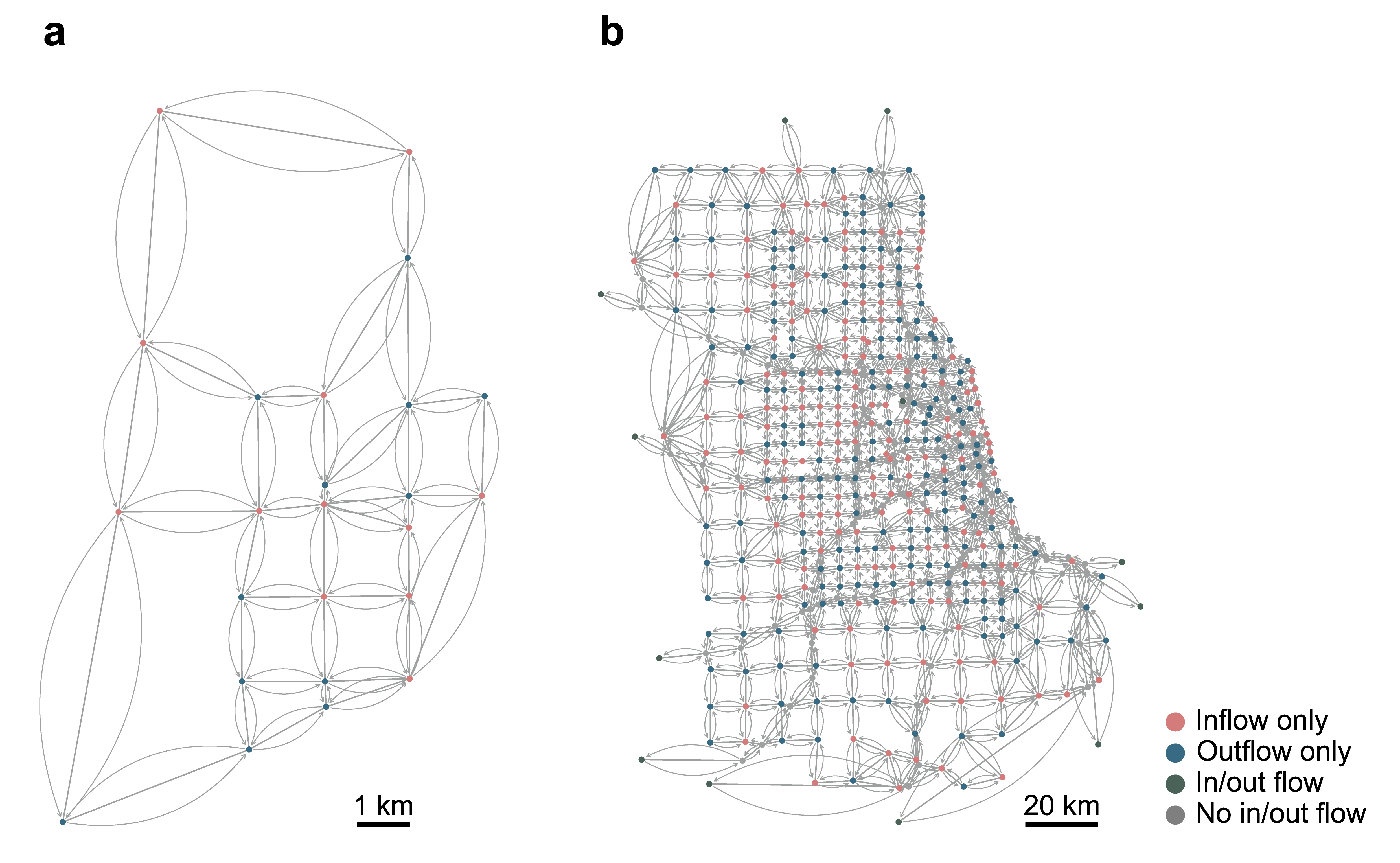}
  \caption{Network structures used in this study.
          (a) Sioux Falls network.
          (b) Chicago Sketch network.
          Different node colors represent different inflow and outflow conditions.
          Bi-directional arrows indicate that there are inbound and outbound links (two links) between two nodes.
          The numbers of nodes/links for the Sioux Falls and Chicago networks are 48/100 and 993/2571, respectively.
   }
   \label{fig:networkfig}
\end{figure}

We used the Sioux Falls and Chicago Sketch networks to verify our model.
Fig.~\ref{fig:networkfig} visualizes these networks with inflow and outflow conditions imposed on the nodes.
We constructed these models based on the network data deposited in the Transportation Networks repository~\cite{tntp}.

%\bibliography{ref}

\begin{thebibliography}{10}

\bibitem{un_2025}
{United Nations}.
\newblock World urbanization prospects 2025: Summary of results.
\newblock Technical Report UN DESA/POP/2025/TR/ NO. 12, United Nations, Department of Economic and Social Affairs, Population Division, New York, 2025.

\bibitem{sweet_2014}
Matthias Sweet.
\newblock Traffic congestion’s economic impacts: Evidence from us metropolitan regions.
\newblock {\em Urban Studies}, 51(10):2088--2110, 2014.

\bibitem{fattah_2022}
Md.~Abdul Fattah, Syed~Riad Morshed, and Abdulla-Al Kafy.
\newblock Insights into the socio-economic impacts of traffic congestion in the port and industrial areas of chittagong city, bangladesh.
\newblock {\em Transportation Engineering}, 9:100122, 2022.

\bibitem{lelieveld_2015}
J.~Lelieveld, J.~S. Evans, M.~Fnais, D.~Giannadaki, and A.~Pozzer.
\newblock {The contribution of outdoor air pollution sources to premature mortality on a global scale}.
\newblock {\em Nature}, 525(7569):367--371, sep 2015.

\bibitem{knittel_2016}
Christopher~R. Knittel, Douglas~L. Miller, and Nicholas~J. Sanders.
\newblock Caution, drivers! children present: Traffic, pollution, and infant health.
\newblock {\em The Review of Economics and Statistics}, 98(2):350--366, 05 2016.

\bibitem{fellendorf_2010}
Martin Fellendorf and Peter Vortisch.
\newblock {\em Microscopic Traffic Flow Simulator VISSIM}, pages 63--93.
\newblock Springer New York, New York, NY, 2010.

\bibitem{horni_2016}
Andreas Horni, Kai Nagel, and Kay~W Axhausen.
\newblock {\em The Multi-Agent Transport Simulation MATSim}.
\newblock Ubiquity Press, London, GBR, 2016.

\bibitem{lopez_2018}
Pablo~Alvarez Lopez, Michael Behrisch, Laura Bieker-Walz, Jakob Erdmann, Yun-Pang Fl^^c3^^b6tter^^c3^^b6d, Robert Hilbrich, Leonhard L^^c3^^bccken, Johannes Rummel, Peter Wagner, and Evamarie Wiessner.
\newblock Microscopic traffic simulation using sumo.
\newblock In {\em 2018 21st International Conference on Intelligent Transportation Systems (ITSC)}, pages 2575--2582, 2018.

\bibitem{barcelo_2010}
Jaume Barcel{\'o}.
\newblock {\em Models, Traffic Models, Simulation, and Traffic Simulation}, pages 1--62.
\newblock Springer New York, New York, NY, 2010.

\bibitem{ma_2002}
Tao Ma and Baher Abdulhai.
\newblock Genetic algorithm-based optimization approach and generic tool for calibrating traffic microscopic simulation parameters.
\newblock {\em Transportation Research Record}, 1800(1):6--15, 2002.

\bibitem{shultz_2004}
Grant~G. Schultz and L.~R. Rilett.
\newblock Analysis of distribution and calibration of car-following sensitivity parameters in microscopic traffic simulation models.
\newblock {\em Transportation Research Record}, 1876(1):41--51, 2004.

\bibitem{park_2005}
Byungkyu~(Brian) Park and Hongtu~(Maggie) Qi.
\newblock Development and evaluation of a procedure for the calibration of simulation models.
\newblock {\em Transportation Research Record}, 1934(1):208--217, 2005.

\bibitem{ma_2007}
Jingtao Ma, Hu~Dong, and H.~Michael Zhang.
\newblock Calibration of microsimulation with heuristic optimization methods.
\newblock {\em Transportation Research Record}, 1999(1):208--217, 2007.

\bibitem{lee_2009}
Jung-Beom Lee and Kaan Ozbay.
\newblock New calibration methodology for microscopic traffic simulation using enhanced simultaneous perturbation stochastic approximation approach.
\newblock {\em Transportation Research Record}, 2124(1):233--240, 2009.

\bibitem{hollander_2008}
Yaron Hollander and Ronghui Liu.
\newblock The principles of calibrating traffic microsimulation models.
\newblock {\em Transportation (Amst.)}, 35(3):347--362, 2008.

\bibitem{zhu_2019}
Li~Zhu, Fei~Richard Yu, Yige Wang, Bin Ning, and Tao Tang.
\newblock Big data analytics in intelligent transportation systems: A survey.
\newblock {\em IEEE Transactions on Intelligent Transportation Systems}, 20(1):383--398, 2019.

\bibitem{baydin_2017}
At\i{}l\i{}m~G\"{u}nes Baydin, Barak~A. Pearlmutter, Alexey~Andreyevich Radul, and Jeffrey~Mark Siskind.
\newblock Automatic differentiation in machine learning: a survey.
\newblock {\em J. Mach. Learn. Res.}, 18(1):5595^^e2^^80^^935637, January 2017.

\bibitem{newbury_2024}
Rhys Newbury, Jack Collins, Kerry He, Jiahe Pan, Ingmar Posner, David Howard, and Akansel Cosgun.
\newblock A review of differentiable simulators.
\newblock {\em IEEE Access}, 12:97581--97604, 2024.

\bibitem{degrave_2019}
Jonas Degrave, Michiel Hermans, Joni Dambre, and Francis wyffels.
\newblock A differentiable physics engine for deep learning in robotics.
\newblock {\em Frontiers in Neurorobotics}, Volume 13 - 2019, 2019.

\bibitem{hu_2020}
Yuanming Hu, Luke Anderson, Tzu-Mao Li, Qi~Sun, Nathan Carr, Jonathan Ragan-Kelley, and Fredo Durand.
\newblock Difftaichi: Differentiable programming for physical simulation.
\newblock In {\em International Conference on Learning Representations}, 2020.

\bibitem{bezgin_2023}
Deniz~A. Bezgin, Aaron~B. Buhendwa, and Nikolaus~A. Adams.
\newblock {JAX-Fluids: A fully-differentiable high-order computational fluid dynamics solver for compressible two-phase flows}.
\newblock {\em Computer Physics Communications}, 282:108527, 2023.

\bibitem{andelfinger_2021}
Philipp Andelfinger.
\newblock Differentiable agent-based simulation for gradient-guided simulation-based optimization.
\newblock SIGSIM-PADS '21, page 27^^e2^^80^^9338, New York, NY, USA, 2021. Association for Computing Machinery.

\bibitem{andelfinger_2023}
Philipp Andelfinger.
\newblock Towards differentiable agent-based simulation.
\newblock {\em ACM Trans. Model. Comput. Simul.}, 32(4), January 2023.

\bibitem{chopra_2023}
Ayush Chopra, Alexander Rodr\'{\i}guez, Jayakumar Subramanian, Arnau Quera-Bofarull, Balaji Krishnamurthy, B.~Aditya Prakash, and Ramesh Raskar.
\newblock Differentiable agent-based epidemiology.
\newblock In {\em Proceedings of the 2023 International Conference on Autonomous Agents and Multiagent Systems}, AAMAS '23, page 1848^^e2^^80^^931857, Richland, SC, 2023. International Foundation for Autonomous Agents and Multiagent Systems.

\bibitem{dyer_2023}
Joel Dyer, Arnau Quera-Bofarull, Ayush Chopra, J.~Doyne Farmer, Anisoara Calinescu, and Michael Wooldridge.
\newblock Gradient-assisted calibration for financial agent-based models.
\newblock In {\em Proceedings of the Fourth ACM International Conference on AI in Finance}, ICAIF '23, page 288^^e2^^80^^93296, New York, NY, USA, 2023. Association for Computing Machinery.

\bibitem{querabofarull_2025}
Arnau Quera-Bofarull, Nicholas Bishop, Joel Dyer, Daniel~Jarne Ornia, Anisoara Calinescu, Doyne Farmer, and Michael Wooldridge.
\newblock Automatic differentiation of agent-based models, 2025.

\bibitem{wu_2018}
Xin Wu, Jifu Guo, Kai Xian, and Xuesong Zhou.
\newblock Hierarchical travel demand estimation using multiple data sources: A forward and backward propagation algorithmic framework on a layered computational graph.
\newblock {\em Transportation Research Part C: Emerging Technologies}, 96:321--346, 2018.

\bibitem{ma_2020}
Wei Ma, Xidong Pi, and Sean Qian.
\newblock Estimating multi-class dynamic origin-destination demand through a forward-backward algorithm on computational graphs.
\newblock {\em Transportation Research Part C: Emerging Technologies}, 119:102747, 2020.

\bibitem{patwary_2023}
A.U.Z Patwary, Shuling Wang, and Hong~K. Lo.
\newblock Iterative backpropagation method for efficient gradient estimation in bilevel network equilibrium optimization problems.
\newblock {\em Transportation Science}, 57(5):1134--1159, 2023.

\bibitem{guarda_2024}
Pablo Guarda, Matthew Battifarano, and Sean Qian.
\newblock Estimating network flow and travel behavior using day-to-day system-level data: A computational graph approach.
\newblock {\em Transportation Research Part C: Emerging Technologies}, 158:104409, 2024.

\bibitem{kim_2024}
Taehooie Kim, Jiawei Lu, Ram~M. Pendyala, and Xuesong~Simon Zhou.
\newblock Computational graph-based mathematical programming reformulation for integrated demand and supply models.
\newblock {\em Transportation Research Part C: Emerging Technologies}, 164:104671, 2024.

\bibitem{zhou_2025}
Xuesong~(Simon) Zhou, Taehooie Kim, Mostafa Ameli, Henan~(Bety) Zhu, Yudai Honma, and Ram~M. Pendyala.
\newblock Flow-through tensors: A unified computational graph architecture for multi-layer transportation network optimization.
\newblock {\em Artificial Intelligence for Transportation}, 1:100006, 2025.

\bibitem{son_2022}
Sanghyun Son, Yi-Ling Qiao, Jason Sewall, and Ming~C. Lin.
\newblock Differentiable hybrid traffic simulation.
\newblock {\em ACM Trans. Graph.}, 41(6), November 2022.

\bibitem{son_2025}
Sanghyun Son, Laura Zheng, Brian Clipp, Connor Greenwell, Sujin Philip, and Ming~C. Lin.
\newblock Gradient-based trajectory optimization with parallelized differentiable traffic simulation.
\newblock In {\em 2025 IEEE International Conference on Robotics and Automation (ICRA)}, pages 14497--14504, 2025.

\bibitem{rios_2013}
Luis~Miguel Rios and Nikolaos~V Sahinidis.
\newblock Derivative-free optimization: a review of algorithms and comparison of software implementations.
\newblock {\em Journal of Global Optimization}, 56(3):1247--1293, 2013.

\bibitem{kusic_2023}
Kre{\v{s}}imir Ku{\v{s}}i{\'{c}}, Ren{\'{e}} Schumann, and Edouard Ivanjko.
\newblock A digital twin in transportation: Real-time synergy of traffic data streams and simulation for virtualizing motorway dynamics.
\newblock {\em Advanced Engineering Informatics}, 55:101858, 2023.

\bibitem{fuller_2020}
Aidan Fuller, Zhong Fan, Charles Day, and Chris Barlow.
\newblock Digital twin: Enabling technologies, challenges and open research.
\newblock {\em IEEE Access}, 8:108952--108971, 2020.

\bibitem{bonabeau_2002}
Eric Bonabeau.
\newblock Agent-based modeling: Methods and techniques for simulating human systems.
\newblock {\em Proceedings of the National Academy of Sciences}, 99(suppl\_3):7280--7287, 2002.

\bibitem{guo_2020}
Ge~Guo and Tianqi Zhang.
\newblock A residual spatio-temporal architecture for travel demand forecasting.
\newblock {\em Transportation Research Part C: Emerging Technologies}, 115:102639, 2020.

\bibitem{li_2016}
Li~Li, Yisheng Lv, and Fei-Yue Wang.
\newblock Traffic signal timing via deep reinforcement learning.
\newblock {\em IEEE/CAA Journal of Automatica Sinica}, 3(3):247--254, 2016.

\bibitem{newell_2002}
G.F. Newell.
\newblock A simplified car-following theory: a lower order model.
\newblock {\em Transportation Research Part B: Methodological}, 36(3):195--205, 2002.

\bibitem{laval_2013}
Jorge~A. Laval and Ludovic Leclercq.
\newblock The hamilton^^e2^^80^^93jacobi partial differential equation and the three representations of traffic flow.
\newblock {\em Transportation Research Part B: Methodological}, 52:17--30, 2013.

\bibitem{seo_2023}
Toru Seo.
\newblock {UXsim}: An open source macroscopic and mesoscopic traffic simulator in {Python} -- a technical overview, 2023.
\newblock arXiv preprint arXiv:2309.17114.

\bibitem{seo_2025}
Toru Seo.
\newblock {UXsim}: lightweight mesoscopic traffic flow simulator in pure {Python}.
\newblock {\em Journal of Open Source Software}, 10(106):7617, 2025.

\bibitem{frostig_2018}
Roy Frostig, Matthew Johnson, and Chris Leary.
\newblock Compiling machine learning programs via high-level tracing.
\newblock In {\em SysML Conference 2018}, 2018.

\bibitem{jax2018github}
James Bradbury, Roy Frostig, Peter Hawkins, Matthew~James Johnson, Chris Leary, Dougal Maclaurin, George Necula, Adam Paszke, Jake Vander{P}las, Skye Wanderman-{M}ilne, and Qiao Zhang.
\newblock {JAX}: composable transformations of {P}ython+{N}um{P}y programs, 2018.
\newblock http://github.com/jax-ml/jax.

\bibitem{train_2003}
Kenneth~E. Train.
\newblock {\em Discrete Choice Methods with Simulation}.
\newblock Cambridge University Press, 2003.

\bibitem{jang_2017}
Eric Jang, Shixiang Gu, and Ben Poole.
\newblock Categorical reparameterization with gumbel-softmax.
\newblock In {\em International Conference on Learning Representations}, 2017.

\bibitem{maddison_2017}
Chris~J. Maddison, Andriy Mnih, and Yee~Whye Teh.
\newblock The concrete distribution: A continuous relaxation of discrete random variables.
\newblock In {\em International Conference on Learning Representations}, 2017.

\bibitem{bengio_2013}
Yoshua Bengio, Nicholas L^^c3^^a9onard, and Aaron Courville.
\newblock Estimating or propagating gradients through stochastic neurons for conditional computation, 2013.
\newblock arXiv preprint arXiv:1308.3432.

\bibitem{margossian_2019}
Charles~C. Margossian.
\newblock A review of automatic differentiation and its efficient implementation.
\newblock {\em WIREs Data Mining and Knowledge Discovery}, 9(4):e1305, 2019.

\bibitem{loshchilov_2018}
Ilya Loshchilov and Frank Hutter.
\newblock Decoupled weight decay regularization.
\newblock In {\em International Conference on Learning Representations}, 2019.

\bibitem{optax}
DeepMind, Igor Babuschkin, Kate Baumli, Alison Bell, Surya Bhupatiraju, Jake Bruce, Peter Buchlovsky, David Budden, Trevor Cai, Aidan Clark, Ivo Danihelka, Antoine Dedieu, Claudio Fantacci, Jonathan Godwin, Chris Jones, Ross Hemsley, Tom Hennigan, Matteo Hessel, Shaobo Hou, Steven Kapturowski, Thomas Keck, Iurii Kemaev, Michael King, Markus Kunesch, Lena Martens, Hamza Merzic, Vladimir Mikulik, Tamara Norman, George Papamakarios, John Quan, Roman Ring, Francisco Ruiz, Alvaro Sanchez, Laurent Sartran, Rosalia Schneider, Eren Sezener, Stephen Spencer, Srivatsan Srinivasan, Milo\v{s} Stanojevi\'{c}, Wojciech Stokowiec, Luyu Wang, Guangyao Zhou, and Fabio Viola.
\newblock The {D}eep{M}ind {JAX} {E}cosystem, 2020.
\newblock http://github.com/google-deepmind.

\bibitem{tntp}
Ben Stabler, Hillel Bar-Gera, and Elizabeth Sall.
\newblock {Transportation Networks}, 2026.
\newblock https://github.com/bstabler/TransportationNetworks.

\end{thebibliography}
%\bibliographystyle{unsrt}

\end{document}